\documentclass[a4paper,traditabstract]{aa}

\usepackage{graphicx}
\usepackage{txfonts}
\usepackage{natbib}
\usepackage{longtable,lscape}
\usepackage{amssymb}
\usepackage{mathrsfs}
\usepackage{amsfonts}
\usepackage{siunitx}
\usepackage{booktabs}
\usepackage{pgf}
\usepackage{multirow}
\usepackage{tabu}
\usepackage{url}
\usepackage{color}
\usepackage[breaklinks, colorlinks, citecolor=blue]{hyperref}
\usepackage{hyphenat}
\usepackage[flushleft]{threeparttable}
\def\Planck{\textit{Planck}}
\def\arcs{\ifmmode {^{\scriptstyle\prime\prime}}
          \else $^{\scriptstyle\prime\prime}$\fi}
\def\parcm{\sa=.08em \sb=.03em
     \ifmmode \hbox{\rlap{.}\kern\sa}^{\scriptstyle\prime}\hbox{\kern-\sb}
     \else \rlap{.}\kern\sa$^{\scriptstyle\prime}$\kern-\sb\fi}
\def\arcm{\ifmmode {^{\scriptstyle\prime}}
          \else $^{\scriptstyle\prime}$\fi}
\def\parcs{\sa=.07em \sb=.03em
     \ifmmode \hbox{\rlap{.}}^{\scriptstyle\prime\kern -\sb\prime}\hbox{\kern -\sa}
     \else \rlap{.}$^{\scriptstyle\prime\kern -\sb\prime}$\kern -\sa\fi}

\def\apjl{ApJL}

\begin{document}

\title{Optical validation and characterisation of \Planck\ PSZ1 sources at the 
Canary Islands observatories. II. Second year of ITP13 observations}
\titlerunning{Optical validation of \Planck\ PSZ1 unknown sources}

\author{R.~Barrena \inst{1,2} \and A.~Ferragamo \inst{1,2} \and J.A.~Rubi\~{n}o-Mart\'{\i}n \inst{1,2} \and 
A.~Streblyanska \inst{1,2} \and A.~Aguado-Barahona \inst{1,2} \and D.~Tramonte \inst{1,2,3}
\and R.T.~G\'enova-Santos \inst{1,2} \and A.~Hempel \inst{4,5} \and H. Lietzen \inst{6} \and N. Aghanim 
\inst{7} \and M. Arnaud \inst{8,9} \and H. B\"ohringer \inst{10} \and G. Chon \inst{10} \and H. Dahle \inst{11} 
\and M. Douspis \inst{7} \and A.~N. Lasenby \inst{12,13} \and P. Mazzotta \inst{14} \and J.B.~Melin \inst{8} 
\and E. Pointecouteau \inst{15,16} \and G.W. Pratt \inst{8,9} \and M. Rossetti \inst{17}}

\institute{Instituto de Astrof\'{\i}sica de Canarias, C/V\'{\i}a L\'{a}ctea s/n, E-38205 La Laguna, Tenerife, Spain\\
\email{rbarrena@iac.es} 
\and
Universidad de La Laguna, Departamento de Astrof\'{i}sica, E-38206 La Laguna, Tenerife, Spain
\and
University of KwaZulu-Natal, Westville Campus, Private Bag X54001, Durban 4000, South Africa
\and
Universidad Andr\'es Bello, Departemento de Ciencias F\'{i}sicas, 7591538 Santiago de Chile, Chile
\and
Max-Planck Institute for Astronomy, K\"onigstuhl 17, D-69117 Heidelberg, Germany
\and
Tartu Observatory, University of Tartu, 61602 T\~oravere, Tartumaa, Estonia
\and 
Institut d'Astrophysique Spatiale, Universit\`e Paris-Sud, CNRS, UMR8617, 91405 Orsay Cedex, France 
\and
IRFU, CEA, Universit\'e Paris-Saclay, F-91191 Gif-sur-Yvette, France
\and
Universit\'e Paris Diderot, AIM, Sorbonne Paris Cit\'e, CEA, CNRS, F-91191 Gif-sur-Yvette, France
\and
Max-Planck-Institut f\"ur extraterrestrische Physik, D-85748 Garching, Germany
\and
Institute of Theoretical Astrophysics, University of Oslo, PO Box 1029, Blindern, 0315 Oslo, Norway
\and
Astrophysics Group, Cavendish Laboratory, JJ Thomson Av., Cambridge, CB3 0HE, UK
\and
Kavli Institute for Cosmology, Madingley Road, Cambridge, CB3 0HA, UK 
\and 
Dipartimento di Fisica, Universit\`a degli Studi di Roma ``Tor Vergata'', via della Ricerca Scientifica, 1, I-00133 Roma, Italy
\and
Universit\'e de Toulouse, UPS-OMP, Institut de Recherche en Astrophysique et Plan\'etologie (IRAP), 31400 Toulouse, France
\and
CNRS, IRAP, 9 avenue Colonel Roche, BP 44346, 31028 Toulouse Cedex 4, France
\and
IASF-Milano, Istituto Nazionale di Astrofisica, via A. Corti 12, 20133 Milano, Italy}

\date{Received ; accepted } 

\authorrunning{Barrena et al.}


\abstract{We report new galaxy clusters previously unknown included in the first \Planck\ 
Sunyaev--Zeldovich (SZ) sources catalogue, the PSZ1. The results here presented have been achieved 
during the second year of a 2-year observational programme, the ITP13, developed at the Roque de los 
Muchachos Observatory (La Palma, Spain). Using the 2.5 m Isaac Newton telescope, the 3.5 m Telescopio 
Nazionale Galileo, the 4.2 m William Herschel telescope and the 10.4 m Gran Telescopio Canarias we 
characterise 75 SZ sources with low SZ significance, SZ S/N$<5.32$. We perform deep optical imaging and 
spectroscopy in order to associate actual galaxy clusters to the SZ \Planck\ source. We adopt robust 
criteria, based on the 2D-spatial distribution, richness and velocity dispersions to confirm actual 
optical counterparts up to $z<0.85$. The selected systems are confirmed only if they are well aligned with 
respect to the PSZ1 coordinate and show high richness and high velocity dispersion. In addition, we also 
inspect the Compton $y-$maps and SZ significance in order to identify unrealistic detections. Following 
this procedure, we identify 26 cluster counterparts associated to the SZ emission, which means that only 
about 35\% of the clusters considered in this low S/N PSZ1 subsample are validated. Forty-nine SZ sources 
($\sim$65\% of this PSZ1 subset) remain unconfirmed. At the end of the ITP13 observational programme, we 
study 256 SZ sources with $Dec \geq -15^{\circ}$ (212 of them completely unknown), finding optical 
counterparts for 152 SZ sources. The ITP13 validation programme has allowed us to update the PSZ1 purity, 
which is now more refined, increasing from 72\% to 83\% in the low SZ S/N regime. Our results are 
consistent with the predicted purity curve for the full PSZ1 catalogue and with the expected fraction 
of false detections caused by the non-Gaussian noise of foreground signals. Indeed, we find a strong 
correlation between the number of unconfirmed sources and the thermal emission of diffuse galactic dust 
at 857 GHz, thus increasing the fraction of false \Planck\ SZ detections at low galactic latitudes.}


\maketitle

\section{Introduction}
\label{sec:intro}
 
Galaxy clusters are extraordinarily useful in studying the structure and evolution of the Universe. The 
$\Lambda$CDM cosmological model makes accurate predictions on the number, abundance and distribution 
of galaxy cluster throughout the Universe, making evident that galaxy clusters play an important role
for constraining cosmological models (e.g., \citealt{evrard89}; \citealt{henry91}; \citealt{White93}; 
\citealt{Eke96}; \citealt{Donahue98}; \citealt{borgani01}). Moreover, cluster abundance studies as a 
function of mass and redshift, N(M,$z$), are powerful cosmological probes (\citealt{Carlstrom2002}; 
\citealt{allen2011}), allowing us to set constraints on parameters such as the dark matter and dark energy 
densities ($\rho_{DM}$, $\rho_{\Lambda}$), or the equation of the state of the dark energy ($w$)
(\citealt{Vikhlinin2009}; \citealt{Mantz2010}).

Throughout the past decades, large-area sky surveys have identified tens of thousands of galaxy clusters by 
using different techniques: such as selecting overdensities of galaxies in optical (e.g., 
\citealt{abell58}; \citealt{abell89}; \citealt{Postman96}; \citealt{wen2009}, \citeyear{WHL2012}) and
infrared (e.g. MaDCoWS \citealt{Gonzalez2019}) surveys, detecting diffuse X-ray sources (e.g. 
\citealt{Ebeling98}; \citealt{Bohringer2000}), tracing the mass gravitional effect produced by the 
weak-lensing footprint (e.g., \citealt{Wittman2006}), and very recently, disentangling the Sunyaev--Zeldovich 
(SZ) effect on the Cosmic Microwave Background (CMB) maps. 

The SZ effect \citep{sz1972} is produced through the inverse Compton scattering when CMB photons interact 
with the high energy electrons of hot intracluster gas, which yields a spectral distortion in the CMB.
This effect has been used to detect galaxy clusters in ground-based milimeter wave surveys, such as
the Atacama Cosmology Telescope (ACT; \citealt{Swetz2011}) and the South Pole Telescope (SPT; 
\citealt{Carlstrom2011}). More recently, the \Planck\footnote{\Planck\ 
(\url{http://www.esa.int/Planck}) is a project of
  the European Space Agency (ESA) with instruments provided by two scientific
  consortia funded by ESA member states and led by principal investigators from
  France and Italy, telescope reflectors provided through a collaboration
  between ESA and a scientific consortium led and funded by Denmark, and
  additional contributions from NASA (USA). } satellite \citep{planck2013-p01}
provided full sky coverage and the opportunity to detect galaxy clusters through the SZ effect 
\citep{planck2011-5.1a,planck2013-p05a,planck2014-a36}. The results of the \Planck\ SZ survey have been
published in two catalogues, the first public SZ catalogue (PSZ1; 
\citealt{planck2013-p05a,planck2013-p05a-addendum}), and the second public SZ catalogue (PSZ2; 
\citealt{planck2014-a36}). 

Using the SZ effect to detect galaxy clusters has the main advantage that the surface brightness 
of SZ signal does not decay with the redshift. So, this allows to detect massive clusters at any 
redshift. The combination of the PSZ1 and PSZ2 catalogues provides 1943 unique SZ detections with 
1330 confirmed clusters and 613 unconfirmed sources. However, in order to make these cluster 
samples useful for cosmology, it is essential to characterise and have an accurate knowledge of 
the \Planck\ SZ sensivity and detection purity. 

In 2010, the \Planck\ collaboration started to confirm SZ cluster candidates cross-correlating with 
previous existing cluster catalogues. The search for possible SZ cluster candidates began with the 
MCXC catalogue \citep{Piffaretti2011} and the AllWISE mid-infrared source catalogue \citep{Cutri2012}. 
Today, many efforts have been done in order to confirm SZ sources in optical wavelength. Example of 
these programmes are that performed using the 
DSS\footnote{DSS: \url{http://stdatu.stsci.edu/dss}} images, the SDSS-DR8 survey 
\citep{sdss-dr8} and the WISE all-sky survey \citep{wise}. In addition, the PSZ1 catalogue was 
cross-correlated to X-ray data, mainly with the ROSAT All Sky Survey \citep{rassbr,rassfaint}, as 
well as other SZ catalogues derived from SPT \citep{Bleem2015} and ACT maps \citep{Marriage2011}. 
Later, several optical follow-up programmes were carried out in large telescopes by obtaining new deep 
images and spectra. This is the case of the programmes performed with the RTT150 telescope 
\citep{planck2014-XXVI}, and the ITP13 (\citealt{planck2016-XXXVI}; \citealt{Barrena2018}, hereafter B+18).

Thus, SZ surveys provide a very valuable tool to construct cluster samples almost complete in 
mass. However, these samples are usefull to determine cosmological parameters and restrict cosmological 
models only if their selection functions are well determined. As SZ surveys only provide flux determinations, 
systematic follow-up programmes of SZ sources are essential in order to carry out the scientific 
exploitation of the resulting catalogues. This is the main goal of this work and, in general, the
main aim of the ITP13 validation programme, so providing a full characterisation of unknown SZ sources
of the PSZ1 catalogue.

This paper is the third part and the last in the framwork of the International Time Project (ITP13), 
a follow-up programme carried out in the Canary Islands Observatories, dedicated to confirm and characterise 
the PSZ1 sources. The ITP13 began with the \citet{planck2016-XXXVI} and followed with B+18. 
This serie of papers includes a detailed study of a complete sample of 212 unknown PSZ1 sources with 
$Dec \geq -15^{\circ}$, which constitutes the ITP13 sample. Here, we present deep photometric and spectroscopic data 
of a sample of 75 SZ sources classifying them as actual clusters counterparts or unconfirmed detections, with 
no optical counterpart. The main aim of the ITP programme is to contribute to improve the
knowledge on the initial \Planck\ SZ completeness, purity and selection function in order to obtain a 
clean PSZ1 sample useful for cosmological studies.

In a similar way, very recently, several optical follow-ups have been focused on PSZ2 sources. Example
of these observational programmes are that performed by \citep{megacam} (hereafter vdB+16) using the MegaCam 
at CFHT, that realized by \citet{alina2018} using SDSS-DR12 database, that carried out in the 4.2m William 
Herschel Telescope by \citet{zohren2019}, or that by \citet{Boada2019} using the 4 m Mayall telescope 
from Kitt Peak National Observatory. Complementary to this ITP13, we have also carried out a second 
optical follw-up (the LP15 long-term observational programme) in order to validate the unknown SZ 
sources of the PSZ2 catalogue using the Roque de los Muchachos Observatory (hereafter ORM) facilities. 
We have characterised 190 PSZ2 sources with Dec$>-15^{\circ}$ studying the purity of this sample. The 
results of these works have been published in \citet{alina2019} and \citet{alejandro2019}.

This paper is organised as follows. Sect.~\ref{sec:sample} describes the observational strategy, and
Sect.~\ref{sec:criteria} includes details on cluster detection technique and validation criteria.
Sect.~\ref{sec:results} provides the cluster counterpart catalogue of SZ sources studied in this work,
as well as, a discussion on the nature of some particular cases. Sect.~\ref{subsec:PSZ1_stat} summarizes 
the whole ITP13 programme providing results on the purity and completeness of the northern PSZ1 sample
and we will expose a possible explanation for a so large number of unconfirmed SZ sources. Finally, 
Sect.~\ref{sec:conclusions} presents the conclusions.

Throughout this paper, we assume a $\Lambda$CDM cosmology, taking H$_0$=100 h km s$^{-1}$ Mpc$^{-1}$,
with h=0.70, $\Omega_m$=0.3 and $\Omega_\Lambda$=0.7.

\section{PSZ1 and observational strategy}
\label{sec:sample}

\subsection{The PSZ1 catalogue and ITP13 subsample}
\label{sec:psz1}

The PSZ1 catalogue (\citealt{planck2013-p05a,planck2013-p05a-addendum}) includes 1227 sources 
selected from SZ effect detections using all-sky maps obtained during the first 15.5 months 
of \Planck\ observations. Very briefly, SZ sources were selected using two Multi-Matched 
Frequency methods (the MMF1 and MMF3) and the PowellSnakes (PwS) technique (see \citealt{Melin2012}) 
with a signal-to-noise ratio (S/N) of 4.5 or higher. The 74\% of the PSZ1 catalogue have been validated, 
determining redshifts for 913 systems, of which 736 are spectroscopic. 753 cluster candidates correspond 
to PSZ1 with Dec$>-15^{\circ}$, which define the PSZ1 reference sample visible from Canary Island 
observatories (hereafter the PSZ1-North sample). 541 of the 753 PSZ1 sources have been validated through 
several follow-ups, while the remaining 212 SZ candidates form the ITP13 sample. \citet{planck2016-XXXVI} 
and B+18 have studied 185 PSZ1 targets, some of them already known but without spectroscopic information. 
In this work, we complete the study of unknown PSZ1 sources characterising the remainning 75 fields. 
The majority of SZ sources characterised during the ITP13 programme show a low S/N in the SZ 
detection, as expected for unknown sources. Only the 12\% present S/N$>6$.

\subsection{Optical follow-up observations}
\label{sec:followup}

The International Time Project ITP13 is a 2-year observational campaign carried out at the 
ORM on the island of La Palma (Spain) from August 2013 to July 2015. B+18 reported the optical 
characterization of 115 SZ sources (with confimed and unconfirmed counterparts) observed during 
the first year of the ITP13 programme. Now, in the present work, we report the optical validation 
carried out in the second and last year, from August 2014 to July 2015. 

Table~\ref{tab:telescopes} lists the telescopes and instruments used to obtain the photometry 
and spectroscopic observations. We use four telescopes: the 2.5 m Isaac Newton Telescope (INT) and 
the 4.2 m William Herschel Telescope (WHT) operated by the Isaac Newton Group of Telescopes (ING),
the 3.5 m Italian Telescopio Nazionale Galileo (TNG) operated by the Fundaci\'on 
Galileo Galilei of the INAF (Istituto Nazionale di Astrofisica), and the 10.4 m Gran 
Telescopio Canarias (GTC) operated by the Instituto de Astrof\'{\i}sica de Canarias 
(IAC). The 2.5 m INT and 4.2 m WHT were used to obtain deep images around SZ \Planck\
pointings, while we perfomed the spectroscopic validation with the 3.5 m TNG and 10.4 m GTC 
telescopes using their respective multi-object spectrographs.

\begin{table*}[h!]
\caption{Telescopes and instruments used in the second year observations of the 
ITP13 programme.}
\label{tab:telescopes}
\begin{center}
\begin{tabular}{c c c S[table-number-alignment = center] c S[table-number-alignment = right] S[table-number-alignment = right]}
\toprule
\multicolumn{1}{l}{Telescope} & {Instrument} & {FoV} & {Pixel Scale $[\arcs]$} & {Resolution} & {N$_{\textrm{ima}}$} &
{N$_{\textrm{spec}}$} \\
\toprule
\multicolumn{1}{l}{2.5 m INT}  &     WFC  & $34\arcm \times 34\arcm$    & 0.33  &   --	  & 38	&     \cr
\multicolumn{1}{l}{3.6 m TNG}  & DOLORES  & $8\farcm6 \times 8\farcm6$  & 0.252 & $R=600$ &	& 10  \cr
\multicolumn{1}{l}{4.2 m WHT}  &    ACAM  & $4\arcmin$ radius           & 0.253 & $R=400$ & 33	&  2  \cr
\multicolumn{1}{l}{10.4 m GTC} &  OSIRIS  & $7\farcm8 \times 7\farcm 8$ & 0.254 & $R=500$ &  1	& 15  \cr
\bottomrule
\end{tabular}
\end{center}
\end{table*}

In order to configure the imaging and spectroscopic sample, we searched for possible SZ counterparts 
in the Sloan Digital Sky Survey (SDSS)\footnote{\url{http://skyserver.sdss.org}} and the Digitized 
Sky Survey (DSS)\footnote{\url{http://archive.stsci.edu/dss}}. The positive identifications of 
this previous screening were directly selected for spectroscopic observations. On the other hand, 
SZ sources with no clear cluster counterparts in the public DSS and SDSS data were included in 
the sample to obtain images with INT and WHT telescope, which allowed a deeper inspection. 

Following the same scheme presented in B+18, and after a previous screening of DSS and SDSS data, 
the observational strategy followed two steps: First, unknown PSZ1 sources were observed by 
obtaining $g'$, $r'$, and $i'$ deep images; then, if SZ counterparts were identified as galaxy 
overdensities, they were definitely confirmed through multi-object spectroscopy (MOS) observations. 
So, cluster were definitely confirmed as actual SZ counterpart and directly linked to the PSZ1 source 
according to the distance to the \Planck\ pointing, richness and velocity dispersion (see 
Sect.~\ref{sec:criteria}).

\subsubsection{Imaging observations, data reduction and photometry}

The High Frequency Instrument (HFI) \Planck\ data provides maps that extend from 100 to 
857\,GHz, with different beam size that varies from $9.6\arcm$ at the lowest frequencies 
to $4.5\arcm$ at the highest. So, the positional error for detecting SZ source using \Planck\
maps is about $2\arcmin$ for targets in the PSZ1 sample \citep{planck2013-p05a}. This result
has been confirmed by comparing \Planck\ SZ and REFLEX II sources. \citet{Bohringer2013} 
found an offset smaller than $2\arcmin$ between X-ray and SZ centres. Even more,
\citet{planck2016-XXXVI} and B+18 found typical offsets of about $2.4\arcm$ and $2.8\arcm$,
respectively, by comparing the positions of optical and SZ centres of PSZ1 sources. On the 
other hand, for PSZ2 counterparts, \citet{alina2018}, \citet{alina2019} and \citet{alejandro2019} 
have also found offsets between $2.6\arcm$ and $4.4\arcm$. Thus, imaging cameras with 
field of view (FOV) larger than $\sim 8\arcmin \times 8\arcmin$ are suitable to prospect the 
sky areas containing \Planck\ SZ counterparts. Therefore, we carried out imaging observations 
using the Wide field Camera (WFC) and the Auxiliary Port Camera (ACAM), respectively, that 
provide a FOV of $34\arcmin \times 34\arcmin$ and $4\arcmin$ radius.

We typicaly acquire 1500\,s and 900\,s exposures in each band using the WFC and ACAM cameras.
A small dithering pattern consisting of 3 frames with $\sim10\arcs$ offset was performed in
order to remove bad pixels, cosmic rays and vignetting zones from the CCD frames. We 
obtain deep imaging in $g'$, $r'$, and $i'$ broad-band filters obtaining
$\sim5-$ and $3-\sigma$ detection levels at magnitudes $r'=21.8$ and 23.2 mag, 
respectively, using the WFC, and $r'=22.0$ and 23.8 mag in the ACAM frames.

The optical WFC and ACAM images were reduced using standard {\tt IRAF} tasks\footnote{{\tt IRAF} 
(\url{http://iraf.noao.edu/}) is distributed by the
  National Optical Astronomy Observatories, which are operated by the
  Association of Universities for Research in Astronomy, Inc., under cooperative
  agreement with the National Science Foundation.}. Raw WFC and ACAM data were corrected from 
bias and flat-field effects. Additionally, the final images were calibrated astrometrically by
using {\tt images.imcoords} IRAF tasks obtaining a precison $\sim 0.1\arcs$ across the vast majority
of the image. Only deviations of $\sim 1\arcs$ are present at the edge of the FOV. The photometric 
calibration was based on SDSS-DR12 standard fields.


We use {\tt SExtractor} \citep{bertin1996} in single-image mode in order to perform the source detection 
and photometry, following the same scheme as in B+18. Objects were detected if they extended more than 
10 pixels showing $S/N>3 \sigma$ in each of the $g'$, $r'$ and $i'$ frames. The {\tt MAGAUTO} procedure 
was applied to obtain photometry in elliptical apertures. Then, the individual $g'$, $r'$ and $i'$ 
photometric catalogues were merged into a master one to include the three-band photomety. This
merging was done by searching for matchings within 1$\arcs$ apertures.

The WFC presents large PSF distortions over the full FOV. This fact made difficult to distinguish
between star and galactic shapes, in particular for sources fainter than $r'=18$ mag. So, we only 
carried out star-cleaning procedures statiscially assuming that cluster regions and the background 
around them show the same distribution of point sources. This step was crucial in order to obtain a
realistic richness and clear red sequences (hereafter RS) in the colour--magnitude diagrams.

All the images obtained in the ITP13 programme have been included in the Virtual Observatory 
collection and are publicly available. In the near future, the photometric and spectroscopic 
catalogues will be also accesible through this web platform.

\subsubsection{Spectroscopic observations and data reduction}
\label{sec:spec_data_subsec}

The majority of the spectra obtained in this observational campaign were acquired using the MOS 
devices of DOLORES and OSIRIS spectrographs at the 3.5\,m TNG and the 10.4\,m GTC, respectively. 
Only a few spectra were obtained using long-slit configuration of OSIRIS, which allow to include 
a few galaxies in each acquisition. The long-slit exposures were acquired using the same instrument
set-up as for the MOS mode. The technical details of these instruments are specified in B+18. Both 
instruments, with a FoV of $\sim 7' \times 6'$, allowed us to sample $\sim 1$ Mpc radius, which 
represent typically about 1 R$_{vir}$ of a cluster at z$\sim 0.3$. For closer clusters, only the 
inner regions are sampled, while for distant clusters we even obtain spectra in their outer 
regions.

As in the first year of observations (B+18), we acquire three exposures of $1800$\,s per mask 
and cluster using DOLORES/TNG and $3\times1000$\,s acquisitions through DOLORES/GTC. These 
integration times, and a typical seeing of $\sim1.1\arcs$, allowed us to obtain $S/N \sim 5$ 
in the spectra of galaxies with magnitudes $r'=20.5$ and 21.5 in the TNG and the GTC,
respectively.

The mask configuration was performed on previous images for each field, placing about 40 
slitlets in DOLORES masks and about 50 slitlets in OSIRIS masks. Galaxy targets (likely cluster
members) were selected from the colour magnitudes obtained with ACAM and WFC photometry. In
addition, we used RGB colour composite images, built from $g'$-, $r'$- and $i'$-band frames, 
as reference. Basically, we select likely cluster members, that means galaxies with coherent 
colours, consistently with the redshift of the cluster and contained within its RS.

The raw spectra were reduced using standard {\tt IRAF} tasks. DOLORES and OSIRIS spectrographs
present a very good and flat spectral response. For this reason, and in order to avoid 
degrading the S/N, bias and flat-field correction were not applied. Only sky subtraction, 
extraction of spectra, wavelength calibration (using He-Ne, Hg, and Ar arcs) and cosmic ray 
rejection procedures were performed. Finally, we checked the OI telluric line ($5577.3$\,$\AA$)
looking for possible deviations. We found small offsets ($\sim 1$\,$\AA$, which mean 
$\sim$50 km\,s$^{-1}$ offset in velocity) in some masks due to flexures of the optical 
system while acquiring data. Spectra were corrected for such effect. 

\begin{figure*}[h!]
\centering
\includegraphics[width=9cm,height=6cm]{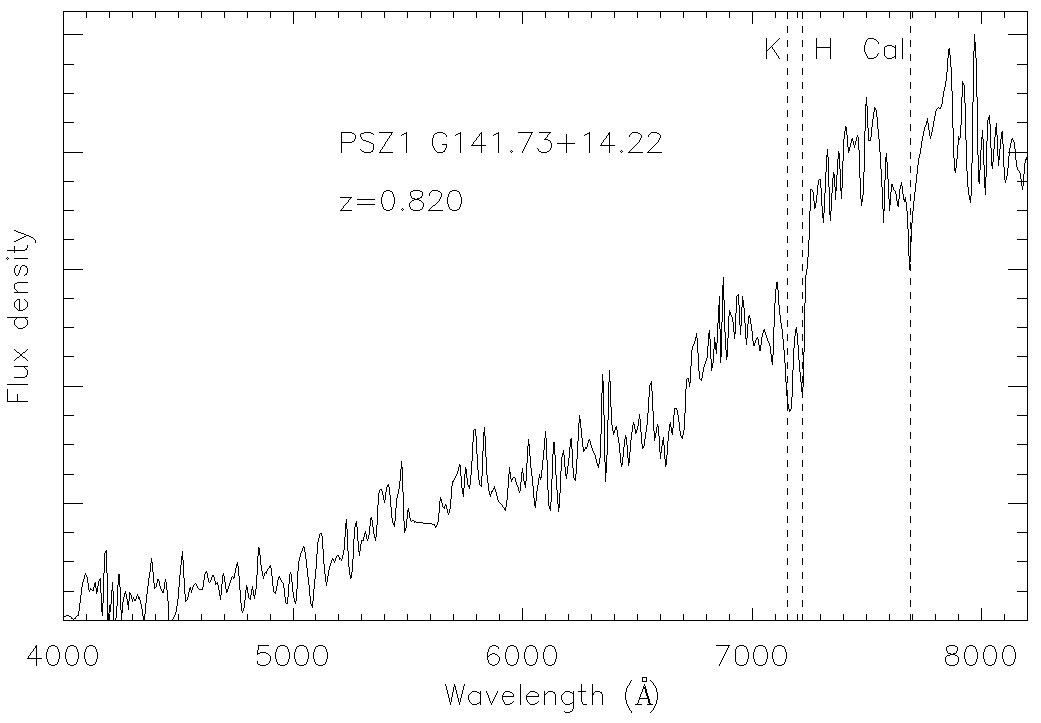}
\includegraphics[width=9cm,height=6cm]{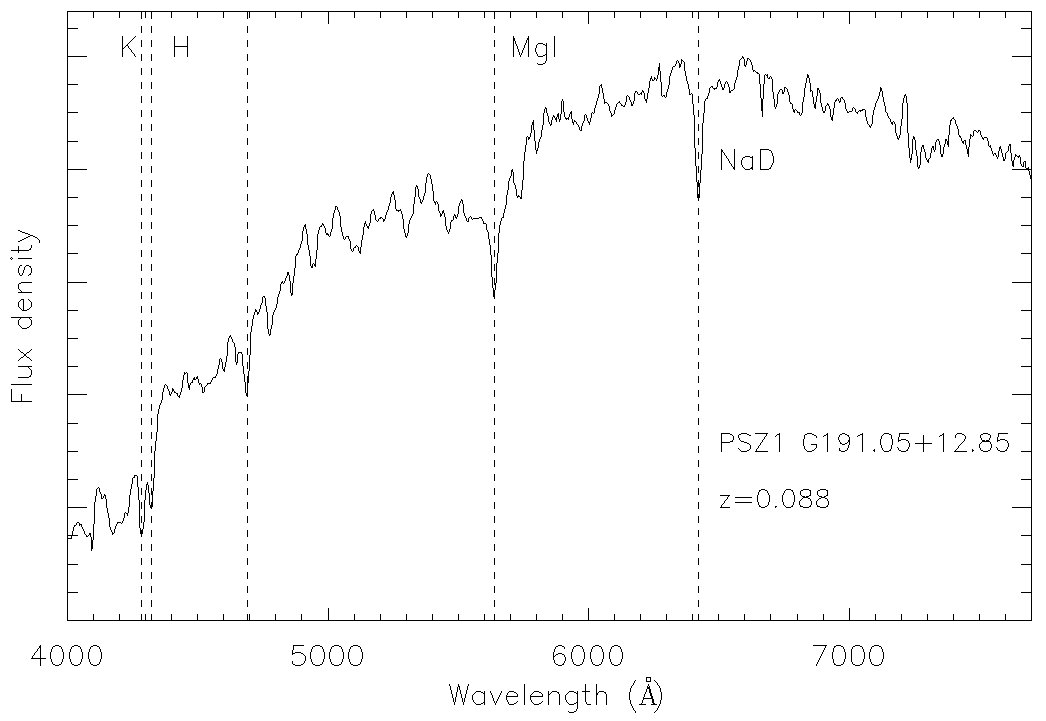}
\caption{Example of the spectra obtained with GTC/OSIRIS (left panel) 
and TNG/DOLORES (right panel) for the two brightest cluster galaxies of G141.73$+$14.22
and G191.05$+$12.85, at $z=0.8208$ and 0.0895, with magnitudes $r'$ = 21.9 and 14.8,
respectively. Vertical dashed lines mark the absorption features identified at the 
redshift of the clusters.}
\label{fig:xc_spec}
\end{figure*}

The final scientific spectra showed $S/N\sim5$ (per pixel around $5000$\,$\AA$) for 
galaxies $r' \sim 20.5$, and $21.7$ mag, observed with the DOLORES/TNG, and OSIRIS/GTC,
respectively. In order to optimize the telescope time, we observed nearby clusters 
($z_{\rm phot} \lesssim 0.35$) in the TNG telescope, while distant clusters (with 
$z_{\rm phot} \gtrsim 0.35$) were included in the target list for the GTC. 
Fig.~\ref{fig:xc_spec} shows two spectra of galaxy cluster members obtained 
with OSIRIS/GTC and DOLORES/TNG.

The redshift determination of galaxies were obtained using the correlation technique
of \citet{tonrydavis79} provided by the {\tt RVSAO/IRAF}\footnote{RVSAO was developed at 
the Smithsonian Astrophysical Observatory Telescope Data Center.} package and the 
{\tt XCSAO} and {\tt EMSAO} tasks. This procedure correlates the scientific spectra 
with several spectrum templates assigning an $R$ parameter associated with the quality 
of the process. We use six spectrum templates of different galaxy morphologies, from 
ellipticals to irregulars, where {\tt XCSAO} and {\tt EMSAO} were able to identify
the main absorptions (H and K doublet, G-band and MgI triplet) and emission features
(mainly OII, OIII doublet, H$_\beta$ and H$_\alpha$ lines). We select the redshift 
estimation corresponding to the highest $R$-value, rejecting that estimates with $R<2.8$,
which show too high errors ($>150$ km\,s$^{-1}$), so they correspond to not reliable 
determinations. Errors in the redshift estimates were $\Delta$v$\sim 80$ km\,s$^{-1}$. 
However, by comparing a set of about 30 spectra with redshifts also available in the SDSS-DR12 
archive, we estimate the systematic errors $\Delta v=110$ km\,s$^{-1}$, which is 
appropriate to sample the velocity dispersion of massive clusters with $\sigma_v$>500 
km\,s$^{-1}$.

As in B+18, we perfom multi-object spectroscopy, which is the ideal technique
to sample the galaxy members in the dense environments of cluster cores. The DOLORES/TNG
and OSIRIS/GTC spectrographs allowed us to obtain up to 45 and 60 redshifts per mask, 
respectively, and so we were able to retrieve 10--25 galaxy members per cluster. The 
success rate\footnote{Percentage of cluster members obtained respect the whole target 
observed in a region.} varies from 80\% to 5\%, with a mean of $\sim$40\%, depending on 
the sampled region (the core or external zones) and the redshift of the cluster. Only 
a very small fraction ($\sim 2$\%) of the targets observed turned out to be stars, mainly 
red objects fainter than $r'=21$.

The galaxy member selection was carried out in two steps: first, we consider galaxy cluster 
members only if they showed radial velocities in the range $\pm 3000$ km\,s$^{-1}$ with respect 
the main value obtained for all galaxies in the cluster; second, we iterate the process 
by selecting galaxies only if they show radial velocities within $\pm$2.7$\sigma_v$ respect 
the mean cluster redshift. This $\sigma - $clipping selection minimises, as much as possible, 
the contamination by interlopers \citep{mamon2010}.


\section{Cluster identification and confirmation criteria}
\label{sec:photoz}

\begin{figure}[ht!]
\centering
\includegraphics[width=8.5cm,height=6cm]{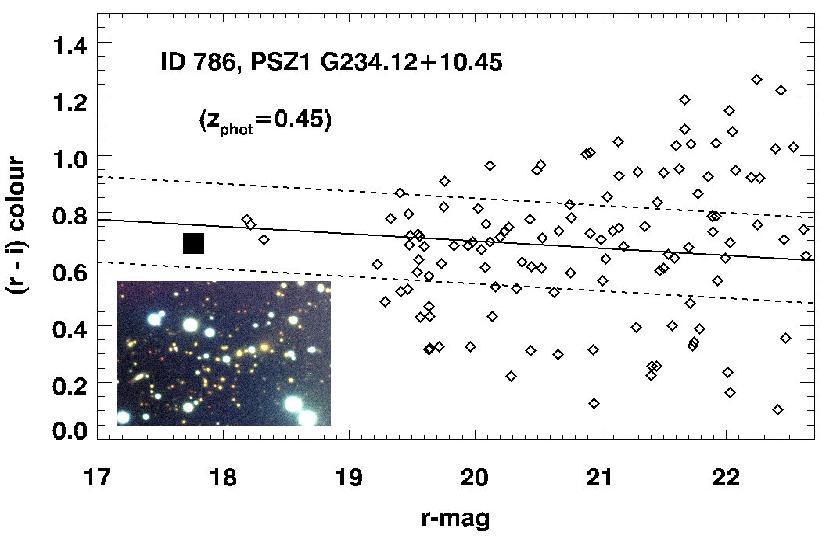}
\caption{($r' - i'$,$r'$) CMD of the source PSZ1 G234.12+10.45. Only
extended sources within a region of $3 \times 2\farcm5$ ($\sim1.3 \times 1.1$ Mpc) around the 
BCG (large filled square) have been considered. The RS (solid line) fit yields a $z_{\rm phot}=0.45$. 
Dashed lines limit the RS$\pm$0.15 mag of likely cluster members. A colour composite image (built 
using the $g'$, $r'$ and $i'$ frames with $3\arcm \times 2.5\arcm$ FOV) is shown 
superimposed to the diagram.}
\label{fig:cmd}
\end{figure}

The main idea behind the cluster identification is to detect overdensities in galaxy 
distributions as the signature of mass concentration. However, given that the galaxy 
population in massive clusters is dominated by early-type morphologies, we look for galaxy 
overdensities showing coherent colours, the RS in the colour-magnitude diagram (CMD; 
\citealt{gladders2000}). This method allows us not only to detect clusters, but also determine 
their photometric redshifts. Following the same prescription detailed in B+18, we selected 
likely galaxy members in the $(g' - r', r')$ and $(r' -i', r')$ CMDs. We identify the brightest 
cluster galaxy (BCG) and we fit the RS considering the five brightest likely cluster members with 
respect to the colour of the BCG in $g' - r'$ or $r' -i'$. Thus, the photometric redshift 
of clusters is derived using the mean colour of likely members (assumed as galaxies with the 
RS$\pm0.15$; see Fig.~\ref{fig:cmd}) and the Eqs. 1 and 2 from \citet{planck2016-XXXVI}. 
Therefore, this process allows us to select cluster member candidates to configure a sample 
of targets for further MOS observations.

We also use RGB colour composite images, constructed by the combination of $g'$-, $r'$-, 
and $i'$- frames to confirm our findings. The galaxy clumps, dominated by a population 
of early-type galaxies, appeared clearly visible in the field. In this way, the visual 
inspection of RGB images proved to be very efficient in cluster identification, in particular for 
clusters at $z>0.7$, where only a few bright galaxies can be detected and the corresponding RS was
not evident. Also, fossil systems \citep[e.g.][]{jones2003,voevodkin2010} are special cases, with
a core dominant (cD) galaxy surrounded by a population of small galaxies, showing a two-magnitude 
gap with respect to the cD. Due to this fact, their RS are not so populated as for normal clusters in 
the bright end, so these systems are difficult to characterise. Therefore, in case of fossil clusters, 
the visual inspection of RGB images turns out to be very useful.

\subsection{Confirmation criteria}
\label{sec:criteria}

The cluster identification is performed following a robust confirmation criteria. We select clusters 
associated to SZ emissions assuming two premises. First, clusters have to be well aligned 
with the SZ peak. Second, clusters showing a SZ signal in \Planck\ maps are expected to be massive systems. 
The first condition is related to the distance between the \Planck\ pointing and the optical centre of 
the clusters. We assume that valid clusters should be placed at $<5\arcmin$, which means 2.5 times the 
typical positional error in the position of \Planck\ SZ detections. This fact is in agreement with the 
findings obtained by \citet{Bohringer2013}, \citet{planck2016-XXXVI} and B+18. The second condition is 
related to the mass of the clusters. Poor systems, groups of galaxies, have not enough mass content 
able to generate a significant SZ signal in \Planck\ maps. 

So, from \citet{planck2014-a30} (see Fig. 1 therein) we assume that \Planck\ SZ clusters should present 
masses M$_{500} \gtrsim 2 \ 10^{14}$M$_\odot \ h_{70}^{-1}$ if they are at $z>0.2$ or 
M$_{500} \gtrsim 10^{14}$M$_\odot \ h_{70}^{-1}$ if they are nearby structures ($z<0.2$). These mass 
values can be interpreted in terms of velocity dispersion, which can be estimated thanks to the 
fact that we have spectroscopic information for a suficient number of cluster members. Therefore, given 
the relation between M$_{500}$ and $\sigma_v$ by \citet{munari2013}, we assume valid SZ counterparts
that clusters at $z<0.2$ showing $\sigma_v > 500$~km\,s$^{-1}$ (M$_{500} > 10^{14}$M$_\odot \ 
h_{70}^{-1}$) and $\sigma_v > 650$~km\,s$^{-1}$ (M$_{500} > 2 \times 10^{14}$M$_\odot \ h_{70}^{-1}$) for
clusters at $z>0.2$.

\begin{table}[h!]
\caption{Confirmation criteria adopted to validate or reject clusters as counterparts of 
SZ detections.}
\label{tab:criteria}
\begin{center}
{\tabulinesep=1.2mm
\begin{tabu}{|c|c|l|c|c|}
\hline
{\tt Flag} & Spectroscopy & $\sigma_v$ limit (km\,s$^{-1}$) & $\sigma_R$  & Dist.  \\ 
\hline
\multirow{2}{*}{1}  & \multirow{2}{*}{YES} & $>500 \ ; \ 0<z<0.2$ & $>1.5$  & $<5\arcmin$  \\  
   &	 & $>650 \ ; \ z>0.2$	& $>1.5$  & $<5\arcmin$  \\  
\hline
2  & NO  & NA			& $>1.5$  & $<5\arcmin$  \\  
\hline
\multirow{3}{*}{3}  & \multirow{2}{*}{YES} & $<500 \ ; \ 0<z<0.2$ & $>1.5$  & $<5\arcmin$  \\  
   &	 & $<650 \ ; \ z>0.2$	& $>1.5$  & $<5\arcmin$  \\  
   & NO  & $-$  		& $<1.5$  & $>5\arcmin$  \\  
\hline
ND & \multicolumn{3}{c}{Non-detection $\equiv$ No galaxy overdensity} & \\
\hline
\end{tabu}}
\end{center}
\end{table}

\begin{figure}[ht!]
\centering
\includegraphics[width=9cm,height=6cm]{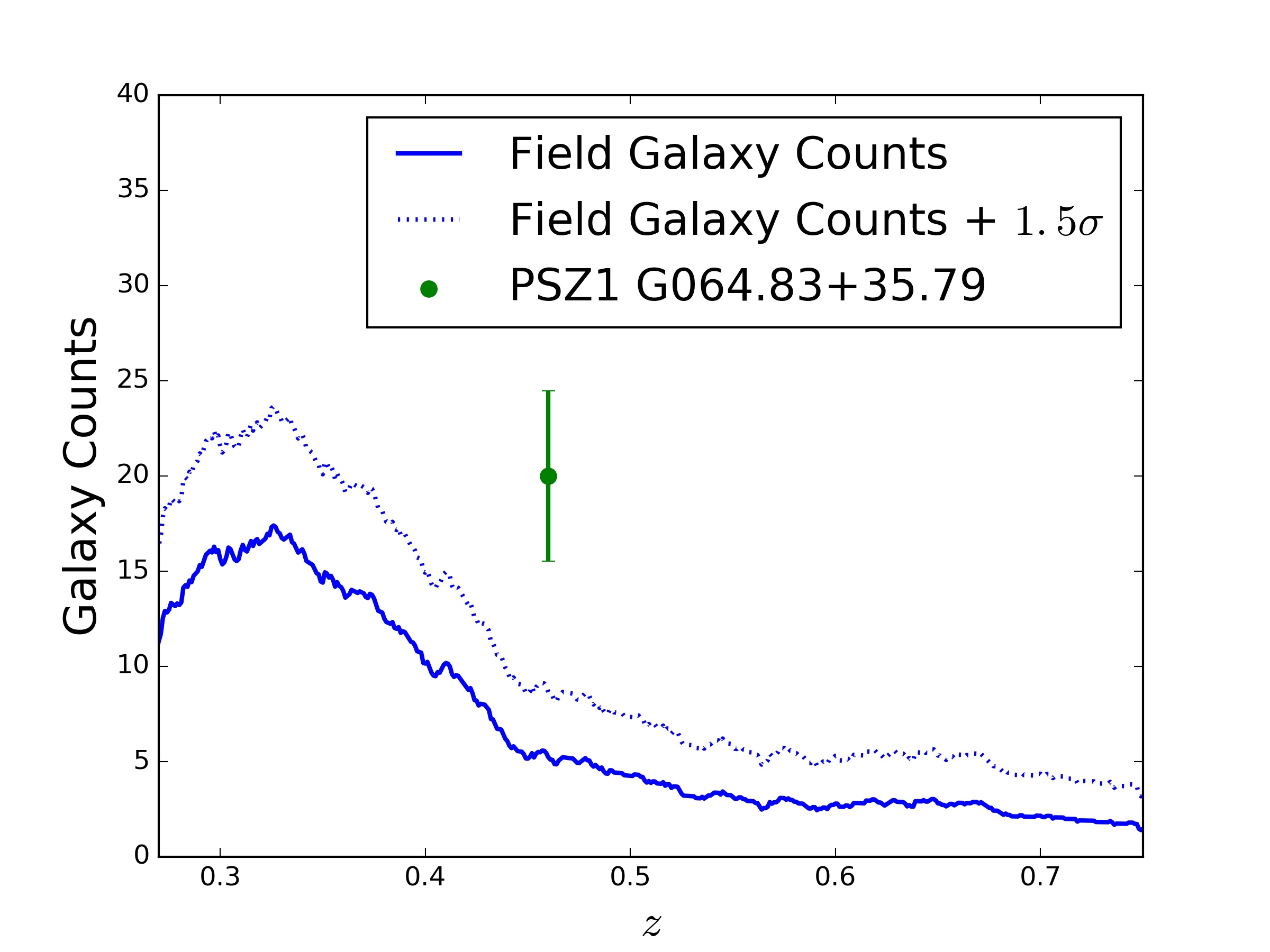}
\caption{Galaxy counts in the field of PSZ1 G064.83+35.79 as a function of the redshift (solid line) 
and its corresponding $+1.5$-$\sigma$ error (dotted line). For this particular case, the richness, 
represented by the green point in the plot, has been computed using the $(r' -i', r')$ 
CMD, so obtaining a $\sigma_R$=6.5 from likely members within 1 Mpc from the cluster center.}
\label{fig:richness}
\end{figure}

However, in cases where we have no spectroscopic information, we confirm cluster candidates using 
the "richness significance". In other words, in the absence of $\sigma_v$ information, galaxy overdensities 
are considered actual SZ counterparts only if they are rich clumps. A detailed description of the method 
here used to compute the richness significance' is shown in \citet{alina2019}. Briefly, for each 
cluster, we count first likely cluster members (assumed as galaxies within the RS$\pm$0.15, in colour) showing 
$r'$-magnitudes in the range $[m_\star - 1$, $m_\star + 1.5$] and laying within 1 Mpc from the 
cluster centre, $R_0$. Then, in order to decontaminate this estimate from the field galaxy contribution, 
we also estimate the richness of the outer regions, $R_f$. Finally, we substract it from the original 
estimate leading to the value $R_{cor} = R_0-R_f$. However, we base our confirmation criterion on the 
significance above the background level $\sigma_R$, which is computed as $R_{cor} / \sqrt R_f$. An 
example of the latter is shown in Fig.~\ref{fig:richness}. Table \ref{tab:criteria} summarizes the 
conditions imposed in order to confirm the clusters as actual SZ counterparts.

For several particular cases, we observe SZ cluster candidates using the ACAM/WHT imaging camera. Given 
that this camera presents a small format FOV, only the inner region of the clusters can be sampled, while 
the galaxy background around that could not be estimated. Therefore, in order to correct the galaxy counts 
in the cluster region ($R_0$), from the field contribution ($R_f$), we make a master background the 
average of the fields observed but excluding a region of 0.5 Mpc radius from those fields where a 
cluster has been detected. These galaxy counts, extracted from the master background, were 
used to compute individual $R_f$ values according to the different redshifts of the clusters, and so to 
obtain individual $\sigma_R$ estimates for the clusters observed with the ACAM/WHT camera.

In the present work, the optical centres of cluster are assumed to be the BCG position. However, in the cases 
showing no evident BCG, the optical centre is supposed to be the centroid (the mean position coordinates)
of likely members. So, we adopt a confirmation criteria based on the distance ($<5\arcmin$), computed
as the projected angle between optical centre and SZ peak emission.

Following the criteria above exposed, based on velocity dispersion, richness significance (as 
mass tracers) and distance to the SZ peak emission, the cluster counterparts are classified in three
different categories, labelled as {\tt Flag}$=\!\!1$, $2$ and $3$, according to their validation level. 
Therefore, clusters classified with {\tt Flag}$=\!\!1$ correspond to massive, rich, and well aligned 
systems confirmed spectroscopically, while clusters classified with {\tt Flag}$=\!\!2$ are also 
considered SZ counterparts, but only photometrically. For such cases, no spectroscopic information 
has been obtained, however, the detections show richness significance ($\sigma_R$> 1.5). Clusters labelled 
as {\tt Flag}$=\!\!2$ may be observed spectroscopically in future works, and so consequently modify 
their {\tt Flag} classification. Clusters classified with {\tt Flag}$=\!\!3$ correspond to galaxy 
systems not well associated to the SZ emission. They could be poor systems, confirmed spectroscopically 
as group-like systems showing low velocity dispersion, or/and systems very distant ($>5'$) from 
the Planck pointing. Thus, we assume these clusters as not valid SZ counterparts. Finally, the 'ND' 
label corresponds to fields where no galaxy overdensity has been detected, that is fields where 
$\sigma_R$ shows no significant peak, $\sigma_R<1.5$.


\section{Results}
\label{sec:results}

Table ~\ref{tab:inpsz1} lists the results for the 75 PSZ1 sources analysed in this work and is organised as 
follows: The first four columns show the identification number in the PSZ1 and PSZ2 lists,
the Planck official name and the significance of the SZ detection reported in the PSZ1 catalogue, 
respectively. Columns 5 and 6 provide the equatorial coordinates (J2000) of the optical centre (assumed 
the BCG position in the most of the cases) of cluster counterparts, while column 7 reports the distance 
between this centre and the SZ one. The double column 8 lists the mean spectroscopic redshift of the 
clusters and that corresponding to the BCG (if available or exist). Column 9 records the number of 
spectroscopic redshifts obtained for cluster members. Column 10 reports the photometric redshift, and 
the multiple column 11 shows the richness and richness significance obtained for each clump. Column
12 provides a {\tt Flag} assignment according to the validation criteria. A final column includes notes
relatives to several cluster counterparts.

In agreement with Sect.~\ref{sec:spec_data_subsec}, the $z_{\rm spec,BCG}$ presents a mean error of 
$\Delta z=0.0004$. However, effects such as the low number of cluster members considered, the presence of
substructures and possible interlopers may produce a mean error in the $<z_{\rm spec}>$ determination of 
$\Delta z \sim 0.001$.

Following the confirmation criteria detailed in the previous section, we find that 26 out of 75 SZ 
sources analyzed present reliable cluster counterparts, 16 of them are definitely confirmed 
spectroscopically (classified with {\tt Flag}$=\!\!1$) and 10 very rich and reliable, with {\tt Flag}$=\!\!2$, 
awaiting for spectroscopic confirmation. On the other hand, almost two thirds of the sample (49 out 
of 75 sources) show no cluster counterparts: 12 sources (with {\tt Flag}$=\!\!3$) are poorly associated 
with their corresponding SZ signal and 37 fields where no galaxy overdensity have been found 
(labelled as `ND'). 

We report spectroscopic information for 29 clusters, 21 of them were observed in MOS mode, using 
DOLORES and OSIRIS spectrographs. The remaining 8 clusters were spectroscopically studied in 
long-slit configuration of ACAM, or were prospected using public data from SDSS-DR12 archive. 
The physical parameters of the clusters here used for the confirmation criteria, such as the 
velocity dispersions and dynamical masses, will be reported and discussed in a future paper 
\citet{ferra2020}.

In the following subsections, we analyze the precision of SZ centres, agreement between 
photometric and spectroscopic redshift, and we discuss on the nature of some clusters showing
particularities, such as fossil clusters or the presence of strong gravitational lensing effects 
(gravitational arcs and arclets; see e.g. \citealt{fort1986}; \citealt{soucail1987}). 


\subsection{Precision of SZ coordinates and redshifts}
\label{distances}

\begin{figure}[ht!]
\centering
\includegraphics[width=6cm,height=6cm]{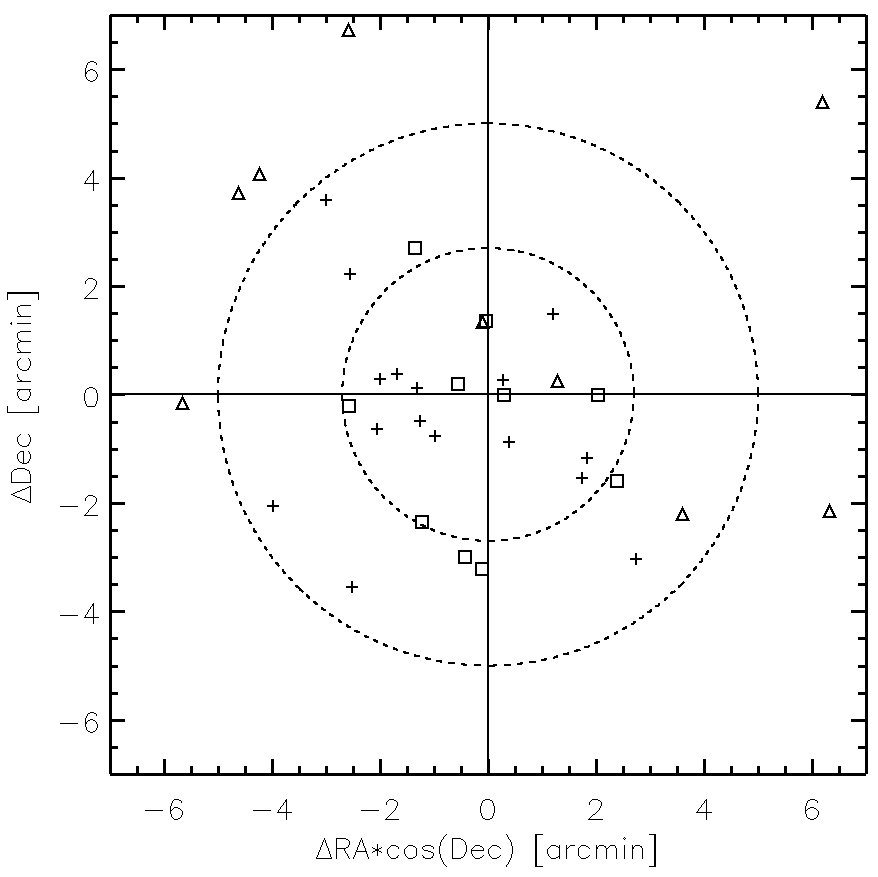}
\caption{Spatial distribution of optical centres with respect their corresponding
PSZ1 coordinates. Crosses, squares and triangles correspond to clusters flagged 
as `1', `2' and `3', respectively. the inner dashed circle marks the $2.8\arcm$ radius
region where 68\% of the confirmed clusters are enclosed. The external dashed 
circle corresponds to the beam size ($5\arcmin$) of SZ \Planck\ detections.}
\label{fig:offsets}
\end{figure}

Fig.~\ref{fig:offsets} shows the spatial distribution of the optical centres of clusters
relative to the PSZ1 coordinates. Ten clusters out of 26 showing actual counterparts present 
centres closer than $2\arcm$ to the PSZ1 coordinate. Moreover, 68\% of the clusters are 
enclosed within $2.8\arcm$, which is in agreement with the finding of $2\arcm$ obtained 
in the REFLEX II X-ray follow-up of \Planck\ SZ sources. 

\begin{figure}[ht!]
\centering
\includegraphics[width=8cm,height=5.5cm]{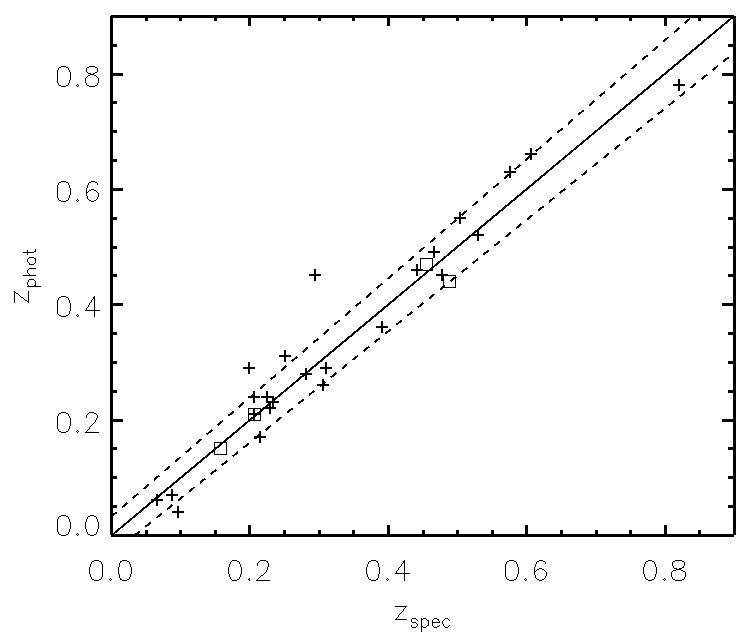}
\caption{Comparison between photometric and spectroscopic redshifts for all the clusters 
characterised in this work and detailed in Table ~\ref{tab:inpsz1}. The solid line
shows the 1:1 relation and dashed lines represent the photometric redshift error 
$\delta z/(1+z) = 0.03$. Crosses represent photometric redshifts obtained using 
our INT and WHT photometry. Squares are redshifts retrieved from SDSS-DR12 archive.}
\label{fig:zphot}
\end{figure}

By comparing the photometric and spectroscopic redshifts of the clusters listed in 
Table~\ref{tab:inpsz1}, we analyse the reliability of the photometric redshift estimates 
(for those clusters with no spectroscopic redshift). Fig.~\ref{fig:zphot} shows this 
comparison, and we find a photometric redshift error of $\delta z/(1+z) = 0.03$, consistent 
with B+18. As we detail above, we obtain photometric redshifts from $(g' - r')$ and $(r' - i')$ 
colours for clusters at low and high redshift, respectively, and as Fig.~\ref{fig:zphot} shows, 
clusters with $z_{phot}<0.45$ and $z_{phot}>0.45$ show the same $rms$ dispersion, $0.033$ in 
both cases. So, both $(g' - r')$ and $(r' - i')$ colours present a good behaviour to determine 
photometric redshifts.

\subsection{Notes on individual clusters}
\label{sec:notes}

In this section we provide a detailed description of the clusters listed in Table 
\ref{tab:inpsz1} showing particular features, such as the presence of gravitational arcs and
fossil systems, or even SZ counterparts reported by other authors or listed in other 
galaxy cluster catalogues. We notice that all images shown in this section present the
north pointing upward and the east to the left.

B+18 found several SZ sources associated with two or more clusters (multiple 
detections). For example, PSZ1 G075.29$+$26.66 includes two clusters, while PSZ1 G125.54$-$56.25 
even encloses three clumps (see B+18). However, in the present work, we do not detect any SZ 
source showing multiple optical counterparts. Only PSZ1 G041.70$+$21.65 (ID-116) and PSZ1 
G326.64$+$54.79 (ID-1140) may present some small galaxy groups very close to the \Planck\ pointing. 
We identify small clumps at (RA=$17\!\!:\!\!47\!\!:\!\!26.63$, Dec=$+17\!\!:\!\!06\!\!:\!\!19.56$) and 
(RA=$13\!\!:\!\!45\!\!:\!\!22.74$, Dec=$-05\!\!:\!\!34\!\!:\!\!41.01$) in the 
ID-116 and ID-1140 fields, respectively. However, none of these galaxy clumps are rich enough
to fit the richness selection criterion, and so not massive enough to be able to produce a SZ signal 
detectable by $Planck$. In agreement with \citet{liu2015} (hereafter Liu+15), PSZ1 G041.70$+$21.65
shows a very massive optical counterpart at z$=0.478$, which we observed spectroscopically
using the OSIRIS/GTC MOS and we estimate a $\sigma_v>1000$~km\,s$^{-1}$. PSZ1 G326.64$+$54.79 
(ID-1140) has also a very rich counterpart, but still has to be observed 
spectroscopically to determine its velocity dispersion.

All the SZ counterparts positively validated and classified with {\tt Flag}$=\!\!1$ have been 
spectroscopically confirmed as clusters with high velocity dispersion. In addition, we have decided 
to confirm the PSZ1 G205.52$-$44.21 (ID-677), validating it with {\tt Flag}$=\!\!1$, although we have only 
one redshift for this cluster. The reason behind this decision is that the ID-677 counterpart is the 
Abell 471 cluster of galaxies, which has been catalogued by \citet{abell89} as a (consolidated) very rich 
system. 

\begin{figure*}
\centering 
\includegraphics[width=183mm,height=70mm]{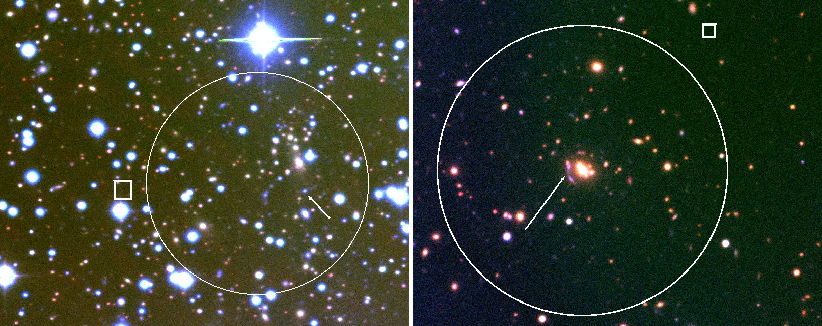}
\caption{RGB images obtained with the WFC/INT (left) and ACAM/WHT (right) $g'$, $r'$ and $i'$ frames 
of the SZ sources PSZ1 G063.92$-$16.75 and PSZ1 G158.34$-$47.49, respectively. The squares mark the 
corresponding SZ \Planck\ coordinates, while the arrows point to the blue gravitational arcs detected 
in the core of the clusters. The circle is a $1.2\arcmin$ radius region including the brightest galaxies 
of each cluster.}
\label{fig:arcs}
\end{figure*}


\paragraph{PSZ1 G063.92$-$16.75 and PSZ1 G158.34$-$47.49}. The SZ cluster counterparts are 
expected to be massive structures, and one observational proof supporting this is the existence 
of strong gravitational effects, such as arcs, in the core of the clusters. The clusters ID-209 
and ID-554 are two examples of SZ counterparts showing gravitational arcs (see Fig.~\ref{fig:arcs}). 
Both clusters have been observed spectroscopically using OSIRIS/GTC MOS, so finding 16 and 8 cluster 
members, respectively. The velocity dispersion estimate confirmes that these two clusters are massive 
systems with $\sigma_v \sim 900$~km\,s$^{-1}$.

\begin{figure}[ht!]
\centering 
\includegraphics[width=90mm,height=60mm]{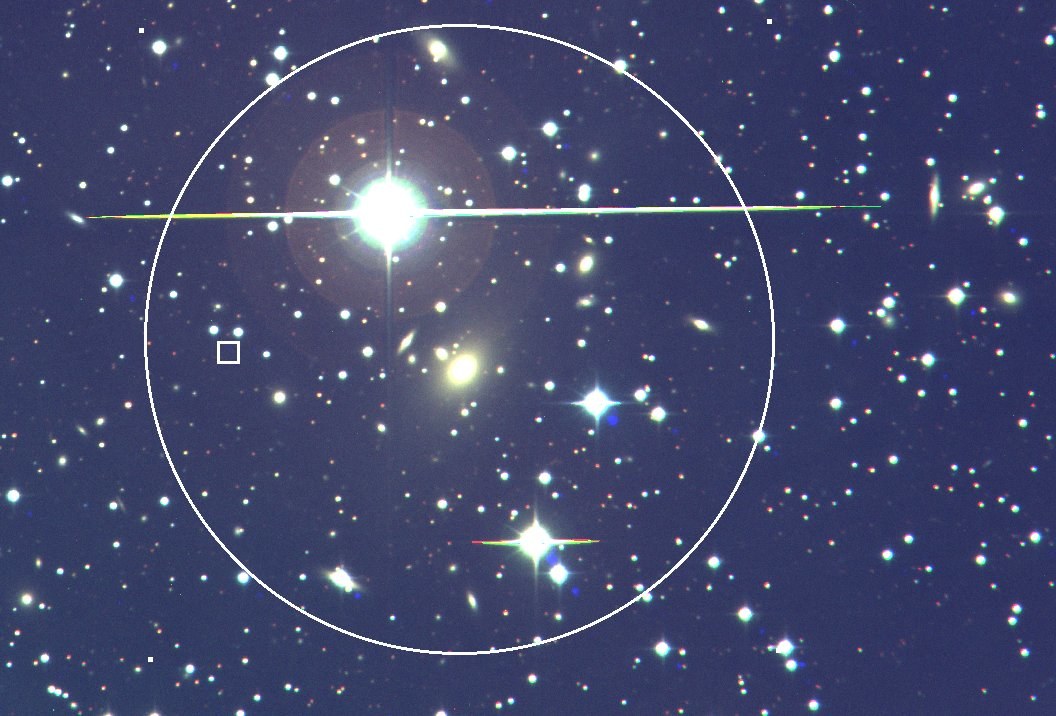}
\caption{Colour composite image obtained with the WFC/INT of the SZ source PSZ1 G097.52$-$14.92.
The central dominant galaxy presents a magnitude $r'<13$ and extends (up to surface brightness 
isophote $\mu_r=25.5$ mag arcsec$^{-2}$) more than $2\arcmin$ ($\sim$100\,kpc at the redshift of the cluster) 
from its centre. The circle encloses a $3\arcmin$ radius region, where the majority of likely 
cluster members, showing $r'>15$ mag are placed. The square marks the \Planck\ SZ peak.}
\label{fig:fossil}
\end{figure}

\paragraph{PSZ1 G097.52$-$14.92 and PSZ1 G249.14$+$28.98}. The ID-340 and ID-846 clusters are
two clear fossil systems, dominated by a huge central galaxy (see Figure \ref{fig:fossil})
and detected with high richness confidence ($\sigma_R \sim 5.8$). Both systems are nearby 
clusters at $z_{phot}=0.04$ and 0.14, respectively. None of this clusters have been observed 
spectroscopically, so we have classified them with {\tt Flag}$=\!\!2$. 

Some of the SZ sources explored in this work, have also been studied by other authors. This is the
case of, for example, PSZ1 G051.42$-$26.16 (ID-162), which has been explored by vdB+16 using 
Megacam at CFHT. In agreement with vdB+16, we do not detect any cluster counterpart associated 
to this SZ source. In a similar way, \citet{planck2014-XXVI} validate the PSZ1 G106.15$+$25.76 
(ID-383) using the RTT150 telescope. Our estimation of $z_{phot}=0.60 \pm 0.04$ is in agreement 
with that obtained by \citet{planck2014-XXVI} spectroscopically, $z_{spec}=0.588$. In addition to 
this source, \citet{planck2014-XXVI} confirms the ID-464 and ID-508. These two SZ sources are 
associated with high redshift clusters at $z=0.577$ and 0.820, respectively. In fact, 
PSZ1 G141.73$+$14.22 constitutes one of the most distant cluster detected by $Planck$. We have 
observed ID-464 and 508 carrying out MOS with OSIRIS/GTC, and we estimate a 
$\sigma_v > 1100$~km\,s$^{-1}$ in both cases.


In agreement with Liu+15, we confirm spectroscopically the cluster at $z=0.478$ as actual counterpart
of the PSZ1 G041.70$+$21.65 (ID-116). We perform MOS with OSIRIS/GTC and, from 25 cluster
members, we estimate a $\sigma_v > 1000$~km\,s$^{-1}$. Regarding the source PSZ1 G306.96$+$50.58
(ID-1080), Liu+15 identify a galaxy overdensity at (RA=$13\!\!:\!\!01\!\!:\!\!46.1$, Dec=$-12\!\!:\!\!04:59$), 
at $7.8\arcm$ to the north of the PSZ1 coordinates, which includes two bright galaxies, quite isolated, at 
$z_{phot}=0.21$. In fact, we find this galaxy clump showing a $\sigma_R<0.5$, so very poor and compatible 
with the number of galaxies of the field. In addition, we also find a richer galaxy overdensity, showing a 
$\sigma_R \sim 3$, but at $\sim 6\arcmin$ from the \Planck\ pointing, compatible with a cluster at 
$z_{phot}=0.58$. However, following our selection criteria, this clump is too far from the \Planck\ pointing 
($>5\arcmin$) to be considered an actual counterpart. Thus we classify this clump with {\tt Flag}$=\!\!3$ 
and consequently, the ID-1080 PSZ1 source remains without a realistic cluster counterpart.

We also cross-correlate our target sample with other optically selected cluster catalogues, such as 
the catalogues by \citet{wen2009} and \citet{WHL2012}. We find one match, which corresponds to
the WHL J35.5193+18.7745, associated with the PSZ1 G150.94$-$39.06 source. This is a very rich cluster
($\sigma_R=9.9$) at $z_{phot}=0.21$ that we classify with {\tt Flag}$=\!\!2$ awaiting for spectroscopic 
validation in the near future.

\subsection{Comments on non-detections}
\label{sec:non-detections}

In our target sample there are several cases of clusters weakly associated with their corresponding 
SZ sources. These are the ID-111, 199, 372, 437 and 646. After prospecting them spectroscopically,
we find small groups showing velocity dispersion in the range $\sigma_v=300-500$~km\,s$^{-1}$,
which are typical for low--mass clusters (M$_{500} \lesssim 10^{14}$M$_\odot \ h_{70}^{-1}$), so
they do not fit our selection criteria. In particular, the ID-646 source seems to be a very clumpy
system (very substructured) at z$=0.24$, and not very rich ($\sigma_R=3$), which shows signs of
being a low-mass system, with a $\sigma_V \sim 300$~km\,s$^{-1}$ (obtained with a few
spectra available in the SDSS-DR12 archive).

In addition, we study in detail two cases, the ID-85 and ID-743, where no clusters, or even poor groups, 
have been identified, and hence, in order to validate them as actual counterparts they have to be 
explored spectroscopically. In the first case, beside the WFC/INT image, we prospect this field 
using GTC/OSIRIS imaging mode, and obtain an image 1.5~$mags$ deeper than the WFC/INT one.
The GTC image reveals that no galaxy overdensity can be associated with this SZ source. For the case 
of ID-743, we perform OSIRIS/GTC MOS observations and we do not identify any actual galaxy concentration 
associated with this SZ source. However, Liu+15 validate this source as a galaxies clump at z$=0.381$ 
that we also detect in the INT/WFC images as a poor group of galaxies with $\sigma_R<1.5$. Consequently, 
we classify this SZ source as `ND' showing none actual cluster counterparts.

Following the validation criteria described in Sect.~\ref{sec:criteria}, we find 49 sources showing no 
galaxy overdensity. This means that {\bf$\sim 65$\%} of the SZ sources analyzed in this work show no optical 
counterpart. In general, the ITP13 has found optical counterparts for the 55\% (140/256) of the SZ sources 
studied, while the 45\% (116/256) remains with not confirmed counterparts. There are three possible reasons 
to explain so many non-detections. First, there could be counterparts at very high redshift ($z>0.85$) that 
can only be detected using near infrared images; the second option is that the \Planck\ SZ detection could be 
affected by the Eddington bias (\citealt{Edington1913}; vdB+16). The third possibility is related to 
the contamination produced by the non-Gaussianity noise of foreground sources.

The detection of clusters at $z>0.85$ using the \Planck\ SZ maps is not very efficient at so high redshift.
In fact, only a few high redshift clusters have been detected (see e.g. \citealt{burenin2018}; vdB+16).
The ITP13 observations are performed using broad-band filters in the visible wavelength range, up to 
$8200$ \AA, so it is true that the detection of galaxy clusters is restricted up to $z<0.85$ in this programme.
However, the detection of a few high redshift clusters does not explain the presence of so many SZ sources 
with not confirmed counterparts. In fact, 116 PSZ1 sources remain unconfirmed (see Sect. 
\ref{subsec:PSZ1_stat}).


On the other hand, the second possibility is that the large fraction of non-detections was related to
the Eddington bias, which is a statistical effect that arises when a threshold in S/N is imposed in a detection
procedure. This may introduce false SZ dectection in the low SZ S/N regime ($\sim 4.5-5.5$), and 
low-mass haloes may be detected. 

\begin{figure}[ht!]
\centering 
\includegraphics[width=\columnwidth]{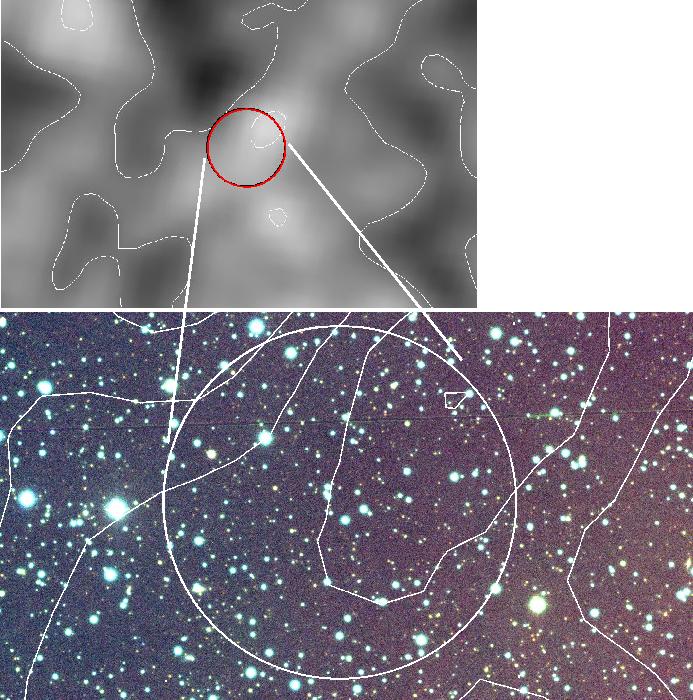}
\caption{\textit{Upper frame:} MILCA Compton $y-$maps in the area around the SZ source PSZ1 
G012.66$+$25.79 (ID-34 showing a 4.5 SZ significance). \textit{Lower frame:} RGB image in the 
region around PSZ1 G012.66$+$25.79. The circle encloses the $5\arcmin$ region centred on the PSZ1 
coordinates of this SZ source. White contours correspond to the SZ emission at $2-$, $3-$ and 
4.5$-\sigma$ detection levels.}
\label{fig:noise}
\end{figure}

A substantial number of SZ sources, mostly detected with very low SZ significance, present very 
irregular and elongated contour shapes of SZ emission. Fig.~\ref{fig:noise} illustrates, as example, 
one of such cases. These cases can be typical candidates affected by Eddington bias. In fact, as 
\citet{alina2018} show (see Fig.~2 therein) how the Compton $y-$maps\footnote{Full-sky Compton parameter 
($y-$map) has been obtained using MILCA procedure \citep{planck2013-p05b} which is publicly available 
since 2015 \citep{PC-XXII}.} of a large fraction of SZ sources classified as non-detection present 
very irregular shapes and no obvious peak at the nominal position of the \Planck\ sources. 

However, the Eddington bias by itself cannot explain the large fraction of SZ sources with not
confirmed counterparts. In fact, the dectection of poor systems (with M$_{500}<10^{14}$M$_\odot 
\ h_{70}^{-1}$) at low redshift ($z<0.3$) is really insignificant or even null.

Regarding the third option, the influence of thermal radio emission of cold galactic dust will
be explored in detail in Sect. \ref{subsec:PSZ1_stat}. We will statistically study the correlation
between high diffuse radio emission zones with not confirmed SZ sources, providing new and detailed 
information on the completeness and purity of the PSZ1 sample.

\section{PSZ1 statistics: purity and contamination}
\label{subsec:PSZ1_stat}

In this section we analyse the statistical properties of the northern PSZ1 sample in order to 
obtain conclusions related to the full PSZ1 catalogue. With the purpose of comparing the PSZ1 
sample with the one observed in the ITP13 programme, we selected PSZ1 sources with 
$Dec \geq -15^{\circ}$, and we name it PSZ1-North. The PSZ1-North sample contains 753 SZ sources 
(61\% of the full PSZ1 catalogue) which is divided in two different subsets. The first 
includes 541 sources already validated at the time of the first PSZ1 publication in 2013 
\citep{planck2013-p05a-addendum} and other optical follow-up programmes, such as that 
performed with the RTT150 telescope \citep{planck2014-XXVI}. We will refer to this as the 
PSZ1-Val sample. The second, called the ITP13 sample, is composed by the 212 
sources that we observed during the ITP13 optical follow-up programme.

\begin{figure}[htbp]
\centering
\includegraphics[width=90mm,height=60mm]{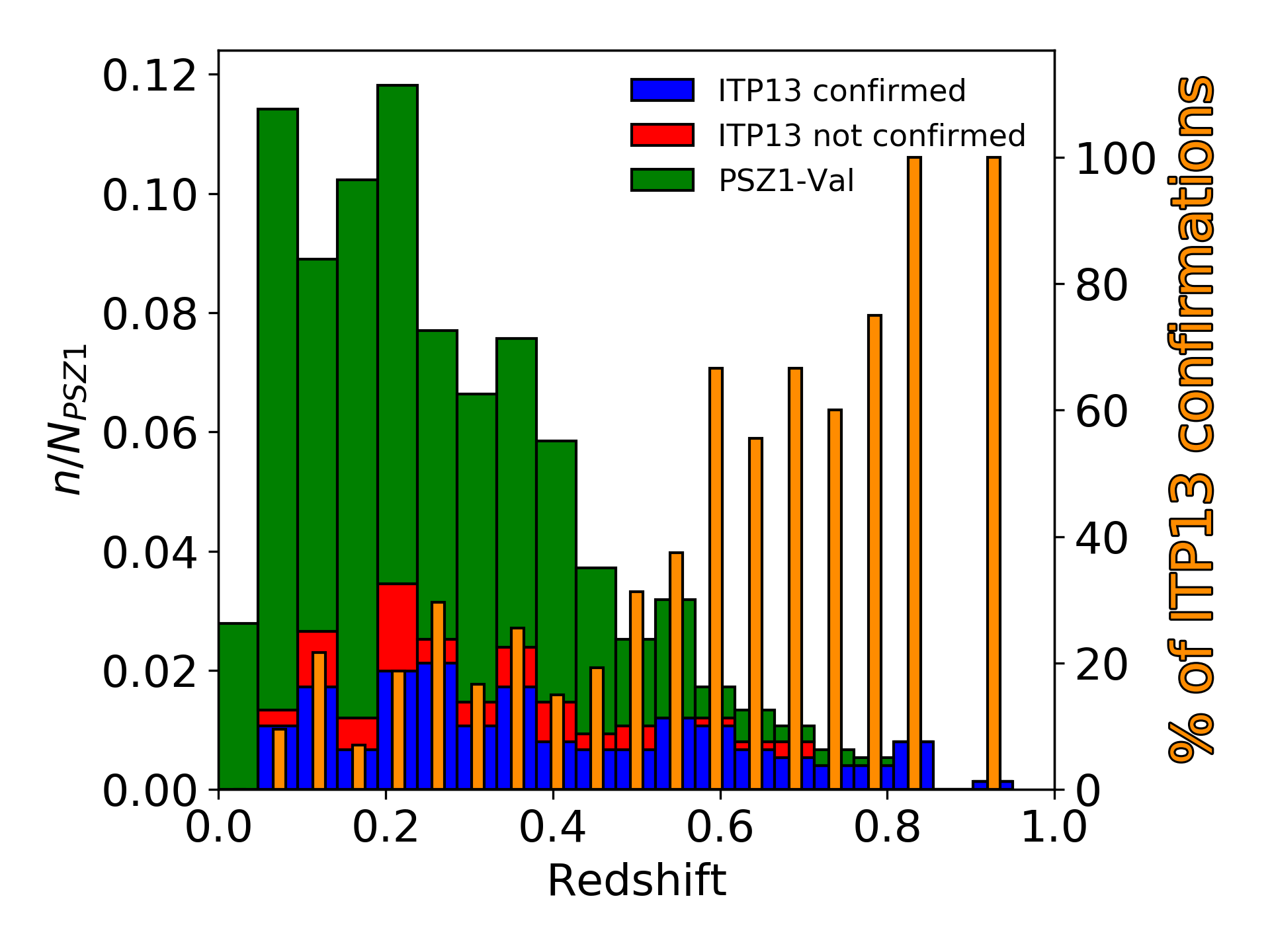}
\caption{Fraction of PSZ1 sources as a function of redshift and normalised 
to the total number of PSZ1-North sources. The PSZ1-Val sample is represented 
in green. Blue and red counts correspond to the ITP sample, classified as confirmed
(with {\tt Flag}=1 and 2) and not confirmed ({\tt Flag}=3) sources, respectively. 
Orange vertical bars show the fraction of SZ sources confirmed during the 
full ITP13 programme.}
\label{fig:PSZ1_red_hist}
\end{figure}

Fig.~\ref{fig:PSZ1_red_hist} shows the number of clusters normalised to the total PSZ1-North 
clusters, and Table~\ref{tab:counts} reports the summary information of the full IT13 programme.
First column specifies the corresponding ITP13 year and reference. Columns 2, 3 and 4 list the 
number of PSZ1 sources observed, confirmed and spectroscopically studied, respectively. The 
remaining columns provide the cluster counts according to the confirmation criteria here adopted.

\begin{table*}[h!]
\caption{Summary information of the full ITP13 programme}
\label{tab:counts}
\begin{center}
\begin{tabular}{l c c c c c c c}
\toprule
{Programme / Reference} & {Observed} & {Confirmed$^*$} & {Spec} & {{\tt Flag}=1} & {{\tt Flag}=2} & {{\tt Flag}=3} & {ND} \\
\midrule
{PC36 $^{**}$}             &  66 &  61 &  42 & $^{***}$ & $^{***}$ & $^{***}$ & 5  \cr
{ITP13 year 1 / B+18}      & 115 &  53 &  56 & 25       & 28       & 13       & 49 \cr
{ITP13 year 2 / this work} &  75 &  26 &  28 & 16       & 10       & 12       & 37 \cr
\midrule
{Total}                    & 256 & 140 & 126 & $>$41  & $>$38 & $>$25 & 91 \cr 
\bottomrule
\end{tabular}
\end{center}

\begin{tablenotes}
\footnotesize
\item $^*$ {\footnotesize Total confirmed clusters correspond to the sum of counts with {\tt Flag}=1 
and {\tt Flag}=2.} 
\item $^{**}$ {\footnotesize PC36 acronism corresponds to \citet{planck2016-XXXVI}, which was developed 
as a previous observational programme.} 
\item $^{***}$ {\footnotesize No flag classification was performed by PC36 and SZ 
sources were classified as "confirmed" and "non-detections". So, total counts in columns 5, 6 and 7 
represent minimum values.}
\end{tablenotes}

\end{table*}

The ITP13 programme has studied a total of 256 sources, 212 of them were completely unknown and 44
had no spectroscopic redshift. At the end, we have positively confirmed 140 optical counterparts and
116 remain unconfirmed. 126 PSZ1 sources at low SZ S/N have been studied spectroscopically 
in the redshift range $0.02<z<0.85$. The ITP13 has confirmed 11 cluster counterparts at $z>0.7$, which 
implies the 80\% of the PSZ1-North sources at this redshift, and all (4) clusters at $z>0.8$. Moreover, 
we have been able to study 30 clusters showing redshifts $z>0.5$, which means the 55\% of the PSZ1-North 
sample at high redshift.

\begin{figure*}[htbp]
\centering
\includegraphics[width=90mm,height=60mm]{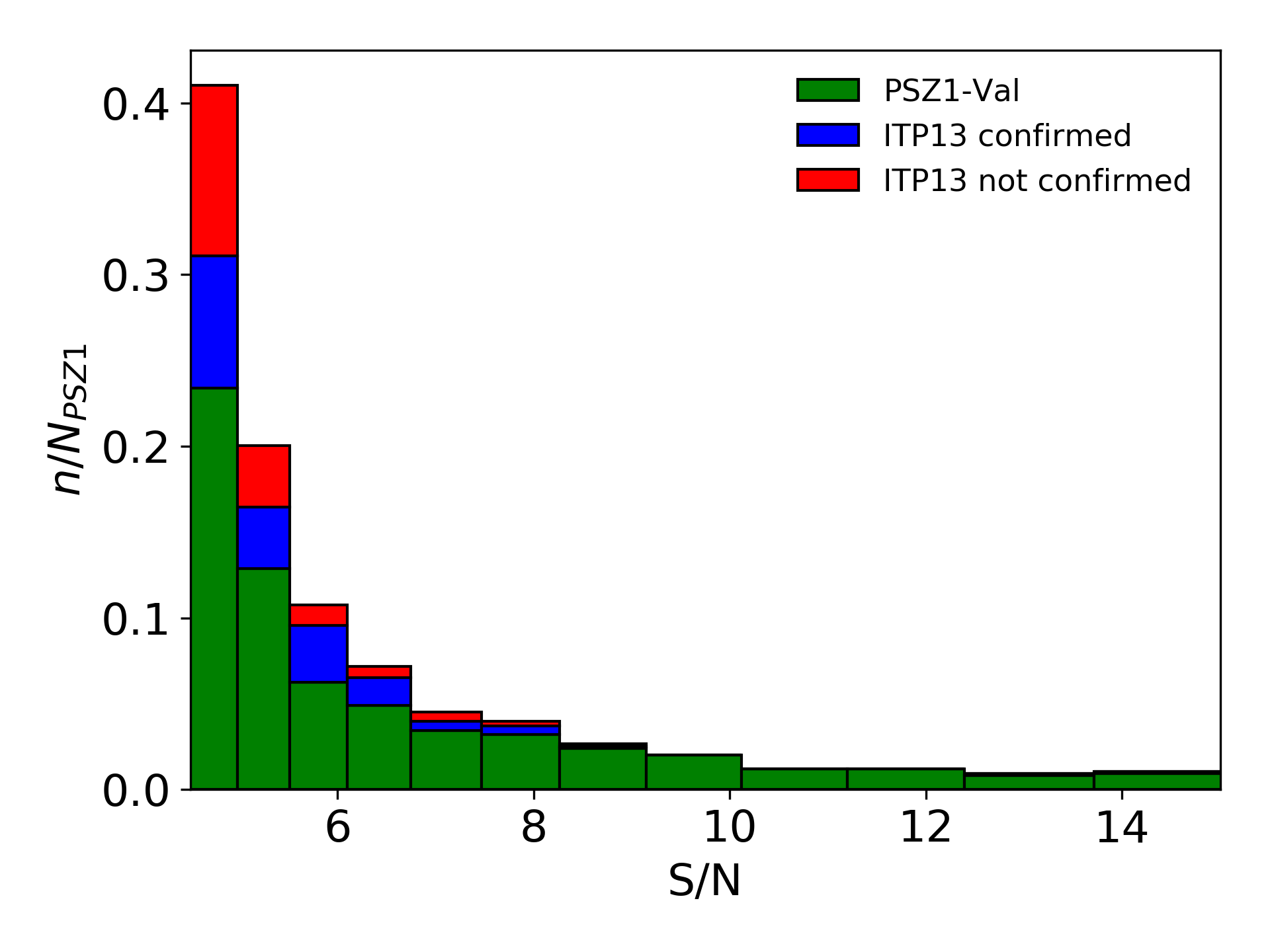}
\includegraphics[width=90mm,height=60mm]{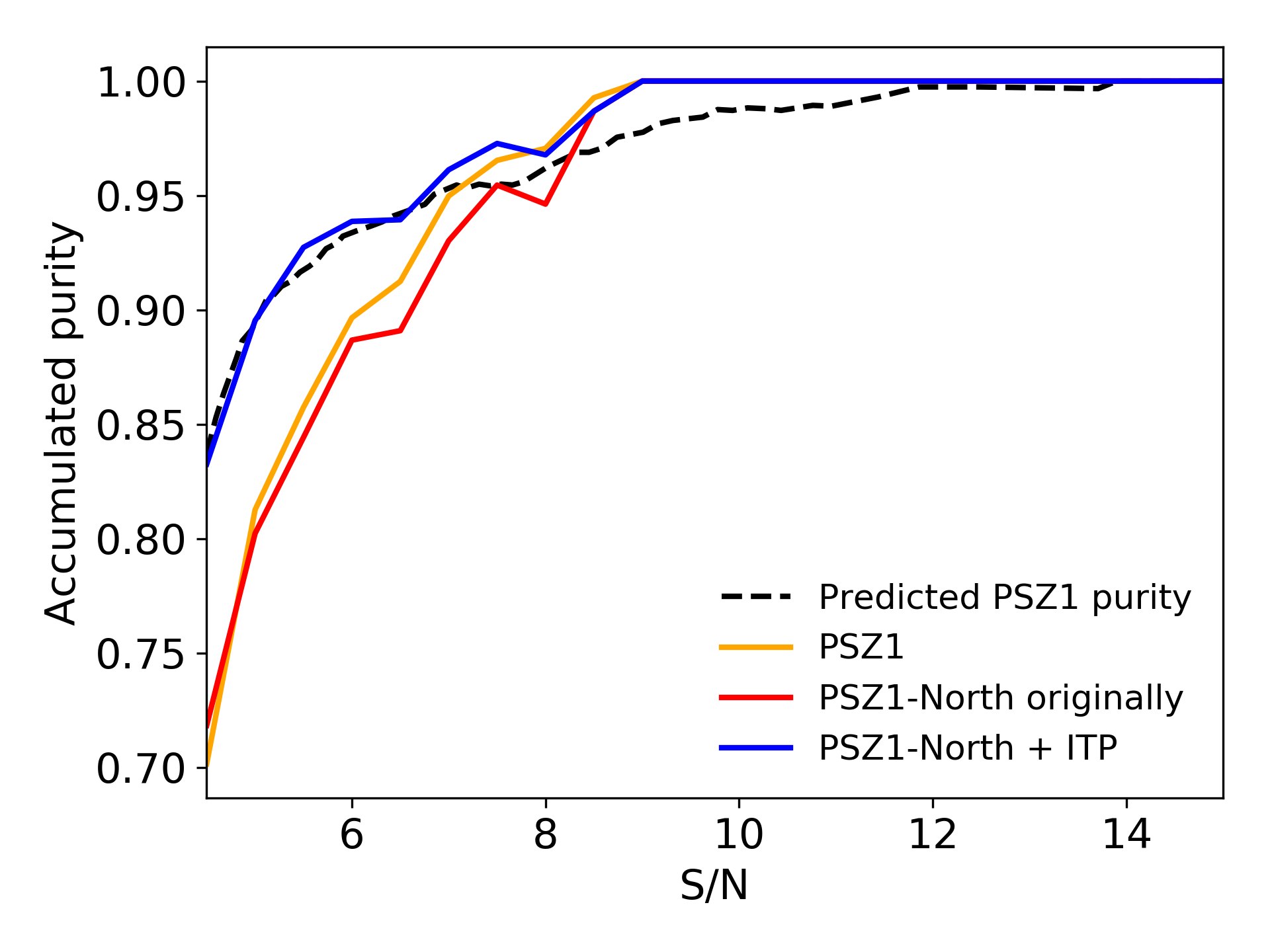}
\caption{\textit{Left panel:} Cluster counts of PSZ1-North sources as function of the SZ S/N
detection. The PSZ1-Val sample is represented in green, which includes SZ validated sources 
in all previous follow-up programmes. The confirmed clusters within the 
ITP13 sample are shown in blue, whereas non validated clusters (classified as {\tt Flag}=3 and ND) 
are shown in red. \textit{Right panel:} Fraction of confirmed clusters as a function of the S/N 
in the PSZ1-North sample. The red line represents the purity at the moment of the publication of 
the PSZ1 catalogue and blue line shows the purity after completing the ITP13 follow-up. The 
orange line represents the accumulated purity of the full PSZ1 catalogue at the moment of 
its first publication, while the black dashed line is the predicted PSZ1 reliability retrived from 
\citet{planck2013-p05a} (see Fig. 10 therein).}
\label{fig:PSZ1_SNR_hist_purity}
\end{figure*}

The left panel in Fig.~\ref{fig:PSZ1_SNR_hist_purity} shows the distribution of PSZ1-North 
sources as function of SZ S/N detection. From the histogram we obtain a median value for
$<S/N>=5.1$, which is comparable to the value that we found for the ITP13 sample ($S/N=4.9$).

We can summarise all this information with accumulated purity. Here, we define the purity as 
the fraction of confirmed clusters with respect to the total number of SZ sources. In 
Fig.~\ref{fig:PSZ1_SNR_hist_purity}, right panel, we show how this purity behaves as function 
of S/N. From this analysis, we conclude that, after all validation programmes, the purity of 
the PSZ1-North sample have passed from 72\% to 83\%, especially at very low S/N due to the
efforts performed in the ITP13 optical follow-up, providing new cluster confirmations.
It is important to remark that the purity of the original PSZ1-North sample (red line) and 
of the full PSZ1 catalogue (orange line) are compatible. This allows us to extrapolate our 
results (in terms of fraction of ND and statistical completeness and purity) to the full PSZ1 
catalogue. So, in this sense, the actual purity of PSZ1 catalogue here obtained (blue curve in 
Fig.~\ref{fig:PSZ1_SNR_hist_purity}) can be qualitatively compared with the simulated one
estimated for the original PSZ1 sample (see Fig. 10. of \citealt{planck2013-p05a}). Both curves
follow a very similar shape and, after completing the ITP13 optical follow-up, we can affirm 
that the PSZ1 presents a $\sim$83\% statistical reliability at S/N=4.5 detection, which is fully 
complatible with the 84\% estimated in \citet{planck2013-p05a}. In addition, we see how the final
PSZ1 reliability is even better than predicted for sources with S/N between 7 and 12, being 100\% 
pure for detections with S/N$>9$. We have to take into account these findings in a qualitative 
way, mainly because many PSZ1 unconfirmed sources in the southern sky still remain without 
any validation.

\begin{figure*}[h!]
\centering
\includegraphics[width=90mm,height=60mm]{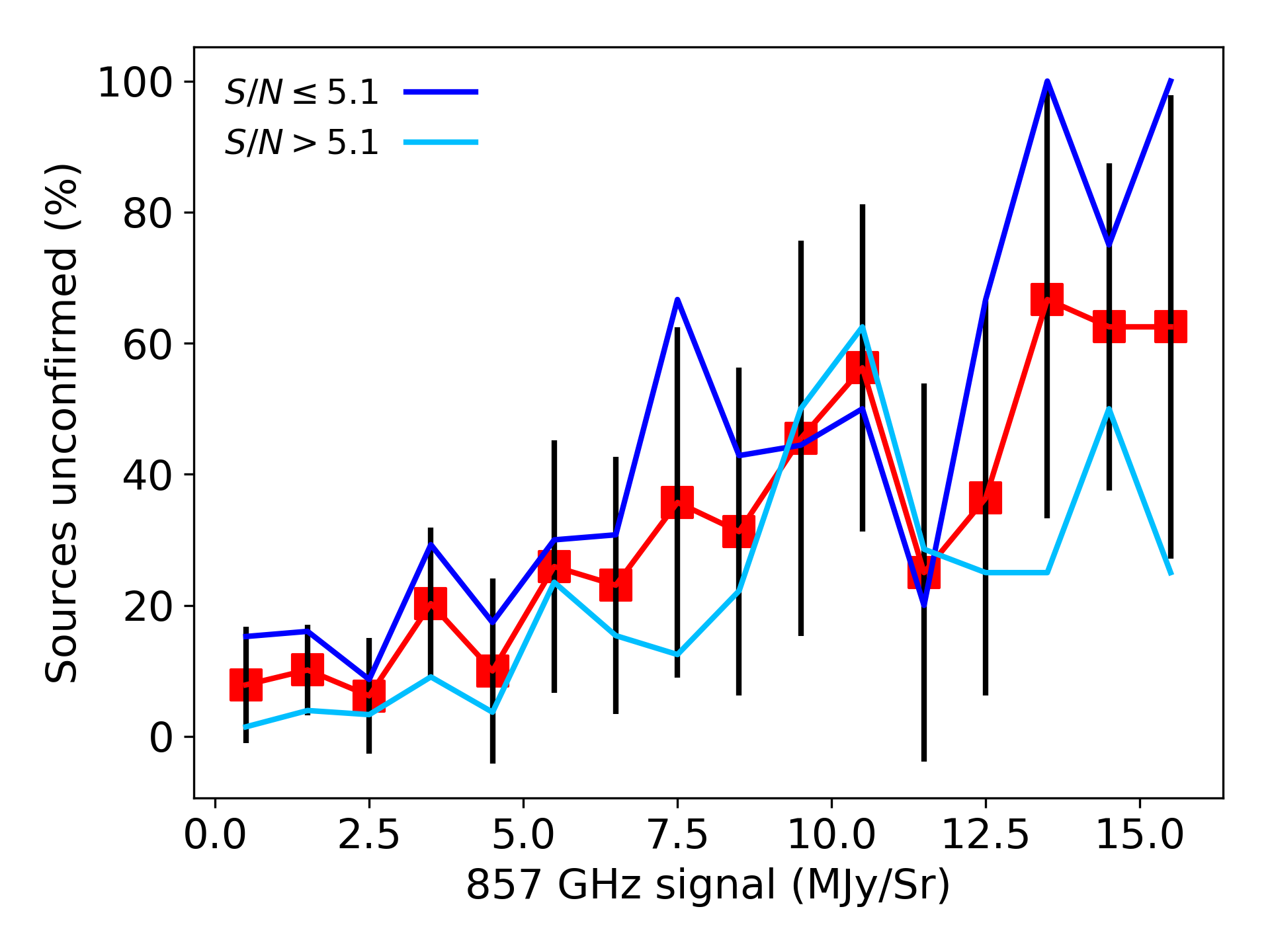}
\includegraphics[width=90mm,height=60mm]{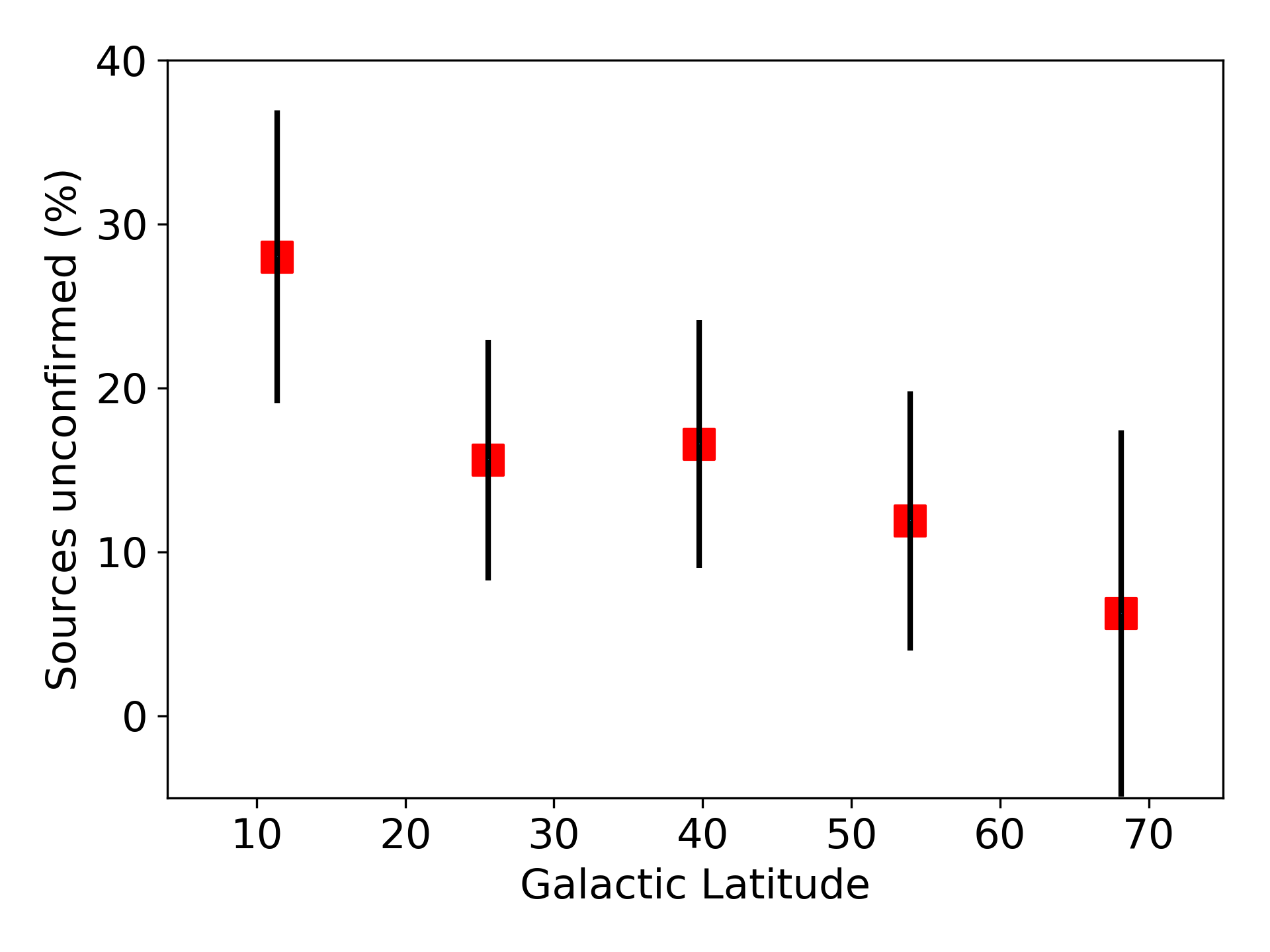}     
\caption{\textit{Left panel:} Fraction of unconfirmed sources with respect the total number of 
elements of the PSZ1-North sample ($Dec\geq-15^{\circ}$) as a function of the 857 \rm{GHz} signal 
in $\rm{MJy\;sr^{-1}}$ taking into account the entire sample (red squares), the $S/N < 5.1$ 
subsample (dark blue line), and the $S/N \geq 5.1$ subsample (light blue line). \textit{Right 
panel:} percentage of unconfirmed sources with respect the total number of elements of the 
PSZ1-North sample as a function of the galactic latitude. Error bars represents the Poisson 
uncertainty of the total number of sources in each bin.}
\label{fig:PSZ1_dust_cont}
\end{figure*}

We noted that the percentage of unconfirmed sources in the low S/N half ($<5.1$) of the 
ITP-North sample is 23.4\%, whereas for $S/N\geq5.1$ this same fraction is only 9.5\%. 
Here, we explore whether these findings could be explained in terms of dust contamination in 
the \Planck\ SZ detections. We studied these sources in detail looking at the radio emission 
in the $857\;\rm{GHz}$ \Planck\ map. In fact, B+18 found hints suggesting that thermal dust 
emission may hardly influence the SZ signal (see Sect.~4.3 therein). This effect may produce 
false SZ detections, especially around regions showing dense galactic dust clouds or even high 
extinction.

Fig.~\ref{fig:PSZ1_dust_cont} shows how the fraction of not validated sources 
({\tt Flag}$=\!\!3$ and ND) with respect to the total number of PSZ1-North sample sources 
(red squares), correlates with the signal at $857\;\rm{GHz}$\footnote{We use here the full-mission 
857GHz Planck maps as downloaded from \url{http://pla.esac.esa.int/pla/}}, computed as the mean signal 
within a region of $0.5^{\circ}$ around the nominal PSZ1 coordinates. In the regions where the 
dust contamination is very low ($<5\; \rm{MJy\;sr^{-1}}$), the fraction of not confirmed 
sources is well below 20\%. However, fields with high dust emission also show high fractions 
($>60\%$) of non-detections. Moreover, taking into account the $S/N\leq 5.1$ (dark blue line) 
and the $S/N>5.1$ (light blue line) bins separately, we note that $\sim100\%$ of low SZ 
S/N sources that show a very high contamination ($>13\; \rm{MJy\;sr^{-1}}$) are non-validated. 

This effect, produced by the cold galactic dust and thus correlated to the galactic latitude (see 
Fig.~\ref{fig:PSZ1_dust_cont}, right panel), has an important effect on the selection function
of the SZ detections using the \Planck\ maps. In particular, we find that the fraction of
unconfirmed sources at low galactic latitudes ($|b|<20^{\circ}$) can be as high as 27\%, while at 
high galactic latitudes ($|b|>60^{\circ}$) is lower than 10\%. We exclude from this analysis, the SZ
sources placed at $|b|>75^{\circ}$ because the low number of targets at this high galactic latitude.
While the number of SZ sources per bin is between 80 and 160 at $|b|<75^{\circ}$, this number counts
drop up to 17 sources at $|b|>75^{\circ}$, making this study not statistically significant (with 
errors larger than 100\%) in this range.

Summarizing, our findings confirm that the fraction of PSZ1 false detections is dominated by 
non-Gaussianity noise of foreground signal, in particular Galactic dust emission, as originaly 
predicted by \citet{planck2013-p05a} (see Fig. 11 therein).

\section{Conclusions}
\label{sec:conclusions}

This article is the third work in the framework of ITP13 PSZ1 validation programme and represents 
the final part of a serie of three papers (see also \citealt{planck2016-XXXVI} and B+18) to carry out
the optical validation of unknown SZ counterparts. The work here presented shows the results obtained 
in the second year of the ITP13 (August 2013 to July 2015) dedicated to characterise unknown PSZ1 
sources in the northern sky (with $Dec \geq -15^{\circ}$) from optical/visible facilities installed 
at the ORM.

We analyze 75 PSZ1 sources by obtaining deep imaging and spectroscopic data, mostly MOS, which allow 
us to study the nature of the detected systems through their 2D galaxy distribution, redshifts and 
velocity dispersion. We assume a robust confirmation criteria (based on the distance to the 
PSZ1 coordinates, richness and velocity dispersion) in order to validate cluster counterparts, and 
label them following a scheme of flags. So, clusters labelled with {\tt Flag}=1
and 2 are confirmed clusters. The first set corresponds to definitive cluster counterparts and 
the second set remains awaiting for velocity dispersion and redshift estimates. False detection
are considered clusters labelled with {\tt Flag}=3 and ND. The former are galaxy overdensities 
very far from the PSZ1 centre or poor systems which do not allow a realistic association with the SZ 
signal. The latter are simply SZ sources with no cluster counterpart. 

Following this classification, we find 26 confirmed counterparts and 49 unconfirmed detections.
So, one third of the sample analyzed in this work show actual cluster counterparts. This low 
fraction of SZ sources with cluster counterparts is mainly due to the low SZ S/N of the
target considered. In fact, we find a clear relation between the fraction of SZ sources catalogued 
as non-detections with the thermal emission at $857\;\rm{GHz}$ produced by the cold galactic 
dust. In particular, we find that almost all the low SZ S/N sources ($\leq5.1$) showing
$>13\; \rm{MJy\;sr^{-1}}$ are non-detections. Thus, this contamination effect statistically correlates 
with the galactic latitude, which confirms that the selection funcion of SZ detections on 
\Planck\ maps displays significant variations over different sky areas. 

We have been able to characterise 212 (validating positively 140 of them) SZ sources, which means 
the 28\% of the PSZ1 sources previously unknown at $Dec \geq -15^{\circ}$. In total, we have provided 
172 and 126 photometric and spectroscopic redshifts of PSZ1 sources, respectively. The contribution 
of the ITP13 programme has been especially important for confirming high redshift cluster 
counterparts. In this sense, the 55\% (80\%) of the clusters with $z>0.5$ ($z>0.7$) have been 
confirmed in the ITP13 framework. The ITP13 results have contributed to characterise the 
17\% of the full PSZ1 sample. 

The results derived from the ITP13 programme have allowed us to obtain a more precise knowledge of 
the PSZ1 purity, confirming that this sample is $\sim$83\% pure for detections at 4.5 SZ significance. 
So, the ITP13 has provided a better understanding of the SZ \Planck\ detections, demonstrating that, 
as originally predicted, false detections are clearly dominated by non-Gaussianity noise of foreground 
signal. 

A similar work has been continued with an additional 2-year observational programme, the LP15, 
dedicated to validate PSZ2 sources in the Northern Hemisphere from Canary Islands observatories 
(see \citealt{alina2019} and \citealt{alejandro2019}).

New observational programmes are planned in order to ultimately confirm additional clusters. In 
addition, in a future work, we will provide the dynamical masses obtained from velocity dispersions 
for the clusters here reported in order to obtain an accurate estimation of the mass bias parameter, 
$(1-b)$, as an attempt to shed some light on the mass scale relation (M$_{SZ}/$M$_{true}$). This 
would allow us to improve the determination of cosmological parameters (mainly $\sigma_8$ and 
$\Omega_m$) using the cluster abundance.

\begin{acknowledgements}

This article is based on observations made with a) the Gran Telescopio Canarias
operated by the Instituto de Astrof\'{\i}sica de Canarias, b) the Isaac 
Newton Telescope, and the William Herschel Telescope operated by the Isaac Newton 
Group of Telescopes, and c) the Italian Telescopio Nazionale Galileo operated 
by the Fundaci\'on Galileo Galilei of the INAF (Istituto Nazionale di Astrofisica).
All these facilities are located at the Spanish del Roque de los Muchachos Observatory 
of the Instituto de Astrof\'{\i}sica de Canarias on the island of La Palma. \\

This research has been carried out with telescope time awarded by the CCI
International Time Programme at the Canary Islands Observatories (programmes
ITP13B-15A).\\

Funding for the Sloan Digital Sky Survey (SDSS) has been provided by the Alfred
P. Sloan Foundation, the Participating Institutions, the National Aeronautics
and Space Administration, the National Science Foundation, the U.S. Department
of Energy, the Japanese Monbukagakusho, and the Max Planck Society.\\

HL is funded by PUT1627 grant from the Estonian Research Council and by the European 
Structural Fundsgrant for the Centre of Excellence "Dark Matter in (Astro)particle 
Physics and Cosmology" TK133. AAB, AF, AS, RB, DT, RGS, and JARM acknowledge financial 
support from Spain's Ministry of Economy and Competitiveness (MINECO) under the 
AYA2014-60438-P and AYA2017-84185-P projects. MR acknowledges financial support from 
contract ASI-INAF n. 2017-14-H.0.

\end{acknowledgements}

\bibliographystyle{aa}

\appendix
 
\section{Cluster validation catalogue obtained in this work}
\begin{landscape}
\begin{table}
\begin{threeparttable}
\caption{Clusters counterparts and PSZ1 sources studied in this work.}
\label{tab:inpsz1}
\small
\begin{tabular}{ccccccccccccl}
\toprule
 \multicolumn{2}{c}{$\mathrm{ID}$} & \Planck\ Name & SZ $S/N$ & R.A. & Decl.& Dist. ($\arcm$) & $<z_{\rm spec}>$~;~$z_{\rm spec,BCG}$&$N_{\rm spec}$&$z_{\rm
phot}$ & $R_c \ \ $;$\ \ \sigma_R$ & {\tt Flag} & Notes \cr
 PSZ1 & PSZ2 & & & \multicolumn{2}{c}{(J2000)} & & & & & & & \cr
\midrule
  32   &      & PSZ1 G012.48$+$27.36 & 4.54 &    $-$      &	  $-$	   & $-$   &	   $-$        & $-$ &  $-$	    & $-$	    & ND & \cr
  34   &      & PSZ1 G012.66$+$25.79 & 4.50 &    $-$      &	  $-$	   & $-$   &	   $-$        & $-$ &  $-$	    & $-$	    & ND & \cr
  38   &      & PSZ1 G015.42$+$58.42 & 4.65 &    $-$      &	  $-$	   & $-$   &	   $-$        & $-$ &  $-$	    & $-$	    & ND & \cr 
  85   &  111 & PSZ1 G031.44$-$19.16 & 4.60 &    $-$      &	  $-$	   & $-$   &	   $-$        & $-$ &  $-$	    & $-$	    & ND & \cr 
 111   &      & PSZ1 G040.17$-$41.51 & 4.57 & 21 32 51.93 & -12 32 33.42 & 5.95  & 0.229 ~;~ 0.2264 &  9  & 0.22$\pm$0.03 & 7.0  ~;~ 1.8  &  3 & \cr 
 112   &      & PSZ1 G040.33$-$16.55 & 4.61 &    $-$      &	  $-$	   & $-$   &	   $-$        & $-$ &  $-$	    & $-$	    & ND & \cr 
 116   &  151 & PSZ1 G041.70$+$21.65 & 4.51 & 17 47 09.18 & +17 11 01.67 & 4.77  & 0.478 ~;~ 0.4773 & 25  & 0.45$\pm$0.05 & 9.0  ~;~ 2.8  &  1 & Liu+15 \cr 
 132   &      & PSZ1 G045.44$-$08.73 & 4.57 &    $-$      &	  $-$	   & $-$   &	   $-$        & $-$ &  $-$	    & $-$	    & ND & \cr 
 135   &  175 & PSZ1 G045.85$+$57.71 & 5.32 & 15 18 20.55 & +29 27 40.50 & 0.39  & 0.607 ~;~ 0.6094 & 37  & 0.66$\pm$0.03 & 39.0 ~;~ 6.2  &  1 & \cr 
 149   &  192 & PSZ1 G048.22$-$51.60 & 4.50 & 22 20 17.44 & -12 11 30.03 & 1.93  & 0.530 ~;~ 0.5322 & 30  & 0.52$\pm$0.02 & 7.5  ~;~ 2.7  &  1 & \cr 
 162   &      & PSZ1 G051.42$-$26.16 & 4.60 &    $-$      &	  $-$	   & $-$   &	   $-$        & $-$ &  $-$	    & $-$	    & ND & vdB+16 \cr 
 193   &      & PSZ1 G058.77$-$26.14 & 4.55 &    $-$      &	  $-$	   & $-$   &	   $-$        & $-$ &  $-$	    & $-$	    & ND & \cr
 199   &      & PSZ1 G059.99$+$11.06 & 4.55 & 19 00 19.37 & +28 58 09.73 & 4.21  & 0.097 ~;~ 0.0974 & 12  & 0.04$\pm$0.03 & 12.0 ~;~ 4.3  &  3 & \cr 
 203   &      & PSZ1 G060.51$-$19.54 & 4.65 & 20 54 31.65 & +13 38 05.57 & 2.04  &	   $-$        & $-$ & 0.59$\pm$0.05 &  9.0 ~;~ 3.0  &  2 & \cr
 209   &  264 & PSZ1 G063.92$-$16.75 & 4.62 & 20 52 51.70 & +17 54 23.02 & 1.73  & 0.392 ~;~ 0.3923 & 16  & 0.36$\pm$0.04 & 46.2 ~;~ 5.8  &  1 & Gravitational arc \cr 
 211   &      & PSZ1 G064.30$+$30.58 & 4.58 &    $-$      &	  $-$	   & $-$   &	   $-$        & $-$ &  $-$	    & $-$	    & ND & \cr
 212   &      & PSZ1 G064.83$+$35.79 & 4.54 & 17 10 08.52 & +40 20 53.64 & 4.35  & 0.442 ~;~ 0.4429 & 22  & 0.46$\pm$0.04 & 14.8 ~;~ 6.5  &  1 & \cr 
 276   &      & PSZ1 G083.35$+$76.41 & 4.65 &    $-$      &	  $-$	   & $-$   &	   $-$        & $-$ &  $-$	    & $-$	    & ND & \cr 
 311   &      & PSZ1 G091.73$-$30.23 & 4.53 &    $-$      &	  $-$	   & $-$   &	   $-$        & $-$ &  $-$	    & $-$	    & ND & \cr 
 332   &  438 & PSZ1 G095.00$-$37.14 & 4.53 &    $-$      &	  $-$	   & $-$   &	   $-$        & $-$ &  $-$	    & $-$	    & ND & \cr
 340   &  456 & PSZ1 G097.52$-$14.92 & 5.06 & 22 37 21.21 & +41 15 01.48 & 2.59  &	   $-$        & $-$ & 0.04$\pm$0.02 & 32.0 ~;~ 5.7  &  2 & Fossil system \cr
 346   &  462 & PSZ1 G098.42$+$77.25 & 4.71 & 13 18 42.87 & +38 43 00.13 & 9.90  & 0.234 ~;~ 0.2348 & 18  & 0.23$\pm$0.02 & 27.1 ~;~ 3.3  &  3 & \cr 
 371   &      & PSZ1 G103.24$+$16.92 & 4.55 &    $-$      &	  $-$	   & $-$   &	   $-$        &     &  $-$	    & $-$	    & ND & \cr 
 372   &      & PSZ1 G103.50$+$31.36 & 4.63 & 17 36 44.28 & +72 34 34.07 & 1.35  & 0.226 ~;~ 0.2257 & 18  & 0.24$\pm$0.04 &  7.4 ~;~ 2.7  &  3 & \cr 
373$^*$&      & PSZ1 G103.56$-$39.35 & 4.50 & 23 47 12.69 & +21 06 02.64 & 5.68  & 0.488 ~;~ 0.4877 &  1  & 0.44$\pm$0.03 &  3.9 ~;~ 1.5  &  3 & \cr 
 383   &  513 & PSZ1 G106.15$+$25.76 & 4.64 & 18 56 51.95 & +74 55 52.73 & 0.58  &	   $-$        & $-$ & 0.60$\pm$0.04 & 10.0 ~;~ 3.2  &  2 & PC26$^+$ \cr 
 398   &      & PSZ1 G109.09$-$52.45 & 4.55 &    $-$      &	  $-$	   & $-$   &	   $-$        & $-$ &  $-$	    & $-$	    & ND & \cr
 430   &      & PSZ1 G117.29$+$13.44 & 4.54 & 23 22 12.90 & +75 19 28.74 & 2.15  & 0.466 ~;~ 0.4677 & 31  & 0.49$\pm$0.04 & 16.0 ~;~ 3.6  &  1 & \cr 
 437   &  579 & PSZ1 G118.87$+$42.71 & 4.51 & 13 35 21.87 & +74 04 10.10 & 1.30  & 0.215 ~;~  $-$   &  8  & 0.17$\pm$0.03 & 12.0 ~;~ 1.4  &  3 & \cr 
 450   &  604 & PSZ1 G123.39$+$30.62 & 4.53 & 12 24 40.09 & +86 27 46.51 & 1.25  & 0.200 ~;~ 0.1979 &  9  & 0.29$\pm$0.03 & 19.6 ~;~ 4.4  &  1 & \cr 
 455   &      & PSZ1 G124.56$+$25.38 & 4.76 &    $-$      &	  $-$	   & $-$   &	   $-$        & $-$ &  $-$	    & $-$	    & ND & \cr
 464   &  630 & PSZ1 G127.02$+$26.21 & 4.95 & 05 58 02.70 & +86 13 49.31 & 1.33  & 0.577 ~;~ 0.5760 & 13  & 0.63$\pm$0.05 & 20.0 ~;~ 5.0  &  1 & \cr 
483$^*$&      & PSZ1 G134.75$-$56.53 & 4.57 & 01 17 21.05 & +05 46 08.09 & 3.39  & 0.206 ~;~ 0.2077 & 11  & 0.21$\pm$0.03 & 20.0 ~;~ 5.0  &  1 & \cr 
 489   &      & PSZ1 G135.68$+$45.61 & 4.58 &    $-$      &	  $-$	   & $-$   &	   $-$        & $-$ &  $-$	    & $-$	    & ND & \cr
 507   &  688 & PSZ1 G141.59$+$23.69 & 4.83 & 06 23 55.16 & +72 50 15.69 & 8.97  & 0.306 ~;~ 0.3071 & 14  & 0.26$\pm$0.03 & 10.9 ~;~ 4.8  &  3 & \cr 
 508   &  689 & PSZ1 G141.73$+$14.22 & 4.97 & 04 41 05.88 & +68 13 15.13 & 0.96  & 0.820 ~;~ 0.8208 & 19  & 0.78$\pm$0.06 & $-$  ~;~ $-$  &  1 & PC26$^+$ \cr 
 515   &  699 & PSZ1 G143.70$-$08.59 & 4.56 &    $-$      &	  $-$	   & $-$   &	  $-$	      & $-$ &  $-$	    & $-$	    & ND & \cr
 517   &      & PSZ1 G144.99$+$54.39 & 4.63 & 11 14 21.86 & +58 23 19.76 & 11.36 & 0.206 ~;~ 0.2063 & 25  & 0.21$\pm$0.02 & 30.0 ~;~ 5.5  &  3 & \cr 
536$^*$&  733 & PSZ1 G150.94$-$39.06 & 5.09 & 02 22 04.63 & +18 46 28.10 & 3.22  &	 $-$	     & $-$ & 0.21$\pm$0.02 & 29.2 ~;~ 9.9  &  2 & WHL J35.5193+18.7745 \cr 
 538   &      & PSZ1 G151.44$-$64.86 & 4.70 & 01 38 13.13 & -04 30 50.78 & 7.20  & 0.158 ~;~ 0.1583 &  2  & 0.15$\pm$0.03 & 11.0 ~;~ 1.8  &  3 & Long-slit spectroscopy \cr 
 554   &  762 & PSZ1 G158.34$-$47.49 & 4.63 & 02 24 56.13 & +08 49 47.59 & 2.17  & 0.311 ~;~ 0.3115 &  8  & 0.29$\pm$0.01 &  7.6 ~;~ 2.8  &  1 & Gravitational arc \cr 
 555   &      & PSZ1 G158.58$-$52.47 & 4.51 &    $-$      &	  $-$	   & $-$   &	  $-$	      & $-$ &  $-$	    & $-$	    & ND & \cr
 557   &      & PSZ1 G159.29$-$24.66 & 4.76 & 03 21 00.12 & +27 26 50.70 & 0.29  &	  $-$	      & $-$ & 0.56$\pm$0.05 & 12.0 ~;~ 3.5  &  2 & \cr
 559   &      & PSZ1 G160.59$+$17.41 & 4.67 & 06 21 11.30 & +53 56 33.81 & 2.87  &	  $-$	      & $-$ & 0.33$\pm$0.04 & 24.0 ~;~ 4.9  &  2 & \cr
 566   &  775 & PSZ1 G163.03$-$27.80 & 4.50 &    $-$      &	  $-$	   & $-$   &	  $-$	      & $-$ &  $-$	    & $-$	    & ND & \cr
 590   &      & PSZ1 G171.01$+$15.93 & 4.89 & 06 35 47.93 & +44 10 15.14 & 1.36  & 0.281 ~;~ 0.2823 & 20  & 0.28$\pm$0.03 & 12.1 ~;~ 3.5  &  1 & \cr 
\midrule
\end{tabular}
\end{threeparttable}
\end{table}
\end{landscape}

\addtocounter{table}{-1}
\begin{landscape}
\begin{table}
\begin{threeparttable}
\caption{Continue. Clusters counterparts and PSZ1 sources studied in this work.}
\label{tab:inpsz1}
\small
\begin{tabular}{ccccccccccccl}
\toprule
 \multicolumn{2}{c}{$\mathrm{ID}$} & \Planck\ Name & SZ $S/N$ & R.A. & Decl.& Dist. ($\arcm$) & $<z_{\rm spec}>$~;~$z_{\rm spec,BCG}$&$N_{\rm spec}$&$z_{\rm
phot}$ & $R_c \ \ $;$\ \ \sigma_R$ & {\tt Flag} & Notes \cr
 PSZ1 & PSZ2 & & & \multicolumn{2}{c}{(J2000)} & & & & & & & \cr
\midrule
 597   &  819 & PSZ1 G173.07$-$36.12 & 4.95 &    $-$      &	  $-$	   & $-$   &	  $-$	      & $-$ &  $-$	    & $-$	    & ND & \cr 
 611   &      & PSZ1 G181.12$-$29.42 & 4.82 &    $-$      &	  $-$	   & $-$   &	  $-$	      & $-$ &  $-$	    & $-$	    & ND & \cr
 625   &  858 & PSZ1 G185.53$-$33.97 & 4.98 &    $-$      &	  $-$	   & $-$   &	  $-$	      & $-$ &  $-$	    & $-$	    & ND & \cr
 629   &      & PSZ1 G186.54$+$85.60 & 4.51 &    $-$      &	  $-$	   & $-$   &	  $-$	      & $-$ &  $-$	    & $-$	    & ND & \cr
 639   &      & PSZ1 G189.29$+$07.44 & 5.10 & 06 34 44.47 & +24 20 42.31 & 2.65  & 0.504 ~;~ 0.5041 &  4  & 0.55$\pm$0.05 & 7.3  ~;~ 1.8  &  2 & Long-slit spectroscopy \cr 
 645   &      & PSZ1 G191.05$+$12.85 & 4.51 & 06 59 50.96 & +25 05 20.20 & 4.10  & 0.088 ~;~ 0.0895 & 15  & 0.07$\pm$0.04 & 36.3 ~;~ 7.1  &  1 & \cr 
 646   &  880 & PSZ1 G191.78$-$26.64 & 4.76 & 04 38 54.25 & +04 49 18.16 & 8.20  & 0.206 ~;~ 0.2082 &  4  & 0.24$\pm$0.04 & 9.0  ~;~ 3.0  &  3 & Long-slit spectroscopy \cr 
 652   &      & PSZ1 G194.70$-$58.93 & 4.62 &    $-$      &	  $-$	   & $-$   &	  $-$	      & $-$ &  $-$	    & $-$	    & ND & \cr
 656   &  897 & PSZ1 G196.62$-$45.50 & 4.72 & 03 42 54.55 & -08 41 08.18 & 3.04  &	  $-$	      & $-$ & 0.25$\pm$0.03 & 88.0 ~;~ 9.4  &  2 & \cr
 659   &      & PSZ1 G198.06$-$12.52 & 4.82 &    $-$      &	  $-$	   & $-$   &	  $-$	      & $-$ &  $-$	    & $-$	    & ND & \cr  
 663   &  906 & PSZ1 G199.70$+$37.01 & 4.89 &    $-$      &	  $-$	   & $-$   &	  $-$	      & $-$ &  $-$	    & $-$	    & ND & \cr
 671   &      & PSZ1 G203.42$-$04.03 & 4.70 &    $-$      &	  $-$	   & $-$   &	  $-$	      & $-$ &  $-$	    & $-$	    & ND & \cr
 677   &      & PSZ1 G205.52$-$44.21 & 4.60 & 03 59 32.89 & -13 39 56.41 & 2.33  & 0.252 ~;~ 0.2523 &  1  & 0.31$\pm$0.03 & 27.0 ~;~ 5.2  &  1 & Abell 471 (L-S spectroscopy) \cr 
 698   &      & PSZ1 G212.63$+$07.45 & 4.72 &    $-$      &	  $-$	   & $-$   &	  $-$	      & $-$ &  $-$	    & $-$	    & ND & \cr
 699   &      & PSZ1 G212.80$+$46.65 & 4.77 &    $-$      &	  $-$	   & $-$   &	  $-$	      & $-$ &  $-$	    & $-$	    & ND & \cr
 743   &      & PSZ1 G223.80$+$58.50 & 4.57 &    $-$      &	  $-$	   & $-$   &	  $-$	      & $-$ &  $-$	    & $-$	    & ND & Liu+15 \cr 
 748   &      & PSZ1 G224.45$+$05.25 & 4.55 & 07 31 13.82 & -07 48 33.48 & 4.47  & 0.067 ~;~ 0.0670 & 17  & 0.06$\pm$0.03 & 41.0 ~;~ 6.4  &  1 & \cr 
 753   &      & PSZ1 G225.39$+$04.23 & 4.52 &    $-$      &	  $-$	   & $-$   &	   $-$        & $-$ &  $-$	    & $-$	    & ND & \cr
 786   & 1065 & PSZ1 G234.12$+$10.45 & 4.61 & 08 09 04.96 & -13 30 29.30 & 2.03  & 0.294 ~;~ 0.2979 & 27  & 0.45$\pm$0.04 & 52.0 ~;~ 7.2  &  1 & \cr 
791$^*$& 1070 & PSZ1 G235.93$+$38.21 & 4.79 & 09 46 34.48 & +00 28 02.13 & 6.66  & 0.455 ~;~ 0.4507 &  2  & 0.47$\pm$0.03 &  5.0 ~;~ 0.8  &  3 & \cr 
 798   &      & PSZ1 G237.94$+$21.65 & 5.05 &    $-$      &	  $-$	   & $-$   &	  $-$	      & $-$ &  $-$	    & $-$	    & ND & \cr
 809   &      & PSZ1 G240.42$+$77.58 & 4.59 &    $-$      &	  $-$	   & $-$   &	  $-$	      & $-$ &  $-$	    & $-$	    & ND & \cr
 846   &      & PSZ1 G249.14$+$28.98 & 4.74 & 09 44 57.58 & -13 48 11.22 & 1.36  &	  $-$	      & $-$ & 0.14$\pm$0.03 & 34.0 ~;~ 5.8  &  2 & Fossil System \cr
 900   &      & PSZ1 G262.45$+$49.34 & 4.60 &    $-$      &	  $-$	   & $-$   &	  $-$	      & $-$ &  $-$	    & $-$	    & ND & \cr
 928   &      & PSZ1 G268.86$+$55.92 & 4.66 &    $-$      &	  $-$	   & $-$   &	  $-$	      & $-$ &  $-$	    & $-$	    & ND & \cr
1080   &      & PSZ1 G306.96$+$50.58 & 4.61 & 13 01 37.25 & -12 08 21.58 & 5.94  &      $-$         & $-$ & 0.58$\pm$0.04 &  6.6 ~;~ 2.9  &  3 & Liu+15 \cr 
1140   & 1539 & PSZ1 G326.64$+$54.79 & 6.61 & 13 45 14.72 & -05 32 04.03 & 3.02  & 	 $-$	     & $-$ & 0.46$\pm$0.05 & 44.0 ~;~ 6.6  &  2 & \cr
1199   &      & PSZ1 G346.57$+$42.71 & 4.60 &    $-$	   & 	 $-$	  & $-$   &	 $-$	     & $-$ &  $-$          & $-$	   & ND & \cr
1217   &      & PSZ1 G355.14$+$55.96 & 4.52 &    $-$	   & 	 $-$	  & $-$   &	 $-$	     & $-$ &  $-$	   & $-$	   & ND & \cr
\bottomrule
\end{tabular}

\begin{tablenotes}
\small
\item $^*$ Studied using SDSS-DR12 photometric and/or spectroscopic data.
\item $^+$ PC26 acronism corresponds to \citet{planck2014-XXVI}
\end{tablenotes}

\end{threeparttable}
\end{table}
\end{landscape}

\end{document}